\documentclass[twocolumn]{aastex631}

\shorttitle{Bicoherence analysis in Swift J1727.8-1613}

\graphicspath{{./}{fig/}}
\begin{document}

\title{The bicoherence analysis of type C quasi-periodic oscillations in Swift J1727.8-1613}

\author[0000-0002-4858-3001]{Haifan Zhu}
\affiliation{Department of Astronomy, School of Physics and Technology, Wuhan University, Wuhan 430072, China}
\author[0000-0003-3901-8403]{Wei Wang}
\altaffiliation{Email address: wangwei2017@whu.edu.cn}
\affiliation{Department of Astronomy, School of Physics and Technology, Wuhan University, Wuhan 430072, China}
\author{Ziyuan Zhu}
\affiliation{Department of Astronomy, School of Physics and Technology, Wuhan University, Wuhan 430072, China}

\begin{abstract}

We present the results of bicoherence analysis for Swift J1727.8-1613 during its 2023 outburst, using data from Insight-HXMT. Our analysis focused on observations with quasi-periodic oscillations (QPOs) of frequencies greater than 1 Hz, revealing that all of them belong to type C QPOs. We found a strong correlation between the QPO frequency and the hardness ratio, as well as a linear relationship between the QPO RMS and the hardness ratio. The bicoherence analysis revealed a transition from a "web" pattern to a "hypotenuse" pattern in the LE and HE energy bands. In the bicoherence patterns, there are correlations between horizontal and vertical bicoherence at $f_1=f_2=f_{\rm QPO}$ with count rates. The diagonal structure at $f_1+f_2=f_{\rm QPO}$ becomes more prominent with increasing energy. Additionally, we discovered a new bicoherence pattern in the medium energy band from 10 -- 20 keV, the diagonal structure at $f_1+f_2=f_{\rm har}$ is prominent only in this energy band, which we refer to as the "parallel"  pattern. The bicoherence analysis indicates that the source is likely a low-inclination source. 
\end{abstract}

\section{Introduction}
\label{sec:intro}
Black hole X-ray binaries (BHXRBs) serve as ideal cosmic laboratories for understanding the physical mechanisms governing accretion dynamics around compact objects. Quasi-periodic oscillations (QPOs), which appear as modest-width peaks in the power density spectra (PDS) of X-ray binaries, are considered excellent probes of the regions closest to these compact objects\citep{van1989quasi}. They can be categorized based on their frequency range into low-frequency QPOs (LFQPOs, 0.1–30 Hz)   and high-frequency QPOs (HFQPOs, $>$100 Hz). LFQPOs, based on their amplitudes, quality factor $Q = \nu / \Delta \nu$ (where $\nu$ represents the frequency of the QPO and  $\Delta \nu$ represents the full width at half maximum; FWHM), and other aspects of the PDS, have been originally classified into three types: A, B, and C. Studying LFQPOs is crucial for enhancing our understanding of the accretion processes around black holes, even though their origin remains a subject of debate
\citep{wijnands1999broadband,casella2005abc,remillard2006x,motta2015geometrical,motta2016quasi,ingram2019}.

Type C QPOs are the most common type of QPOs, with a central frequency range of approximately 0.01-30 Hz \citep{casella2005abc}. They are characterized by high RMS amplitudes of up to 20\% and exhibit strong band-limited noise, indicating that the variability in X-ray brightness is concentrated within a certain frequency range. Additionally, they often display second and sub-harmonics, meaning there are oscillations at twice and half the frequency of the main QPO \citep{ingram2019}.

Various models have been proposed to explain the origin of LFQPOs, centering on either the geometric configuration of the accretion disk or its intrinsic characteristics, and these models broadly encompass mechanisms based on instabilities and geometrical effects.
Geometric models often consider Lense–Thirring precession, suggesting that the oscillations are caused by the precession of the inner accretion disc due to the frame-dragging effect of a spinning black hole \citep{stella1997lense,stella1999correlations,schnittman2006precessing}. A variant precession model suggests that type C QPOs arise from the precession of a finite inner region of the accretion flow, rather than just a particle or small ring \citep{ingram2009low,ingram2011physical}. 
\cite{ma2021discovery} have discovered LFQPOs at energies exceeding 200 keV in MAXI J1820+070, suggesting these oscillations likely originate from the precession of a small-scale jet, challenging most existing models with their significant soft lag and distinctive energy-dependent behaviors. In a more detailed analysis, \cite{ma2023detailed} have suggested that the high-energy LFQPOs originate from the Lense-Thirring precession of the relativistic jet, while the low-energy radiation is primarily emanating from the perpendicular innermost regions of the accretion disk. 

Intrinsic models delve into harmonic oscillations inherent to specific characteristics of the accretion flow. For instance, the accretion ejection instability model \citep{Tagger1999,varniere2002accretion}, illustrates a spiral wave instability occurring within the density and scale height of a slender disk, permeated by a robust vertical (poloidal) magnetic field. 
Another model, the propagating oscillatory shock model, posits interactions between the disk and corona as its fundamental premise \citep{chakrabarti1993smoothed, Molteni1996, chakrabarti2008evolution}.

Although various models successfully replicate the frequencies of the peaks in PDS, there is no consensus on the exact mechanism that produces these QPOs. Hence, it is essential to go beyond the PDS analysis and apply more advanced signal analysis methods. Applying wavelet analysis to the study of MAXI J1535–571 reveals a connection between the phenomenon of QPOs and the corona \citep{chen2022wavelet,chen2022waveletB}. Applying wavelet analysis to the Neutron Star
Interior Composition Explorer (NICER) observations of MAXI J1535-571, \cite{chen2024different} uncovers noteworthy distinctions in PDS, hardness ratio, and mean count between light curve segments above and below the wavelet confidence level. This analysis also reveals distinct S-factor (defined as $S = \frac{\tau_{\text{eff}}}{\tau_{\text{sel}}}$,where $\tau_{\text{eff}}$ and $\tau_{\text{sel}}$ are the effective oscillation time duration and the
total lengths, respectively, of the selected time ranges in seconds) values for type C and type B QPOs, providing a novel approach for distinguishing between the two types of QPOs. 

The Hilbert–Huang Transform (HHT) is a method utilized for analyzing signals that are both nonlinear and nonstationary \citep{Huang1998}. 
\cite{yu2023hilbert} applied the Hilbert–Huang Transform to analyze the data of MAXI J1535-571, and their comparison with Fourier analysis results suggests that the broadening of the QPO peak mainly arises from frequency modulation. Further examination indicates a potential shared physical origin between these modulations and broadband noise, consistent with the internal shock model of the jet. \cite{shui2024recovery} also applied this method to MAXI J1535-571, and the results indicate a strong correlation between the QPOs originating from the jet and corona, supporting the scenario of jet–corona coupling precession. These studies indicate that the HHT method is an effective tool for investigating QPOs, thereby improving our understanding of their origin. 

\cite{maccarone2002higher} propose that bicoherence, an indicator of phase coupling at different Fourier frequencies, is a valuable tool for studying higher-order variability. \cite{maccarone2011coupling} discovered that observations exhibiting type C QPOs generally manifest three distinct patterns: the "hypotenuse," "cross," and "web". Conversely, the behavior of type B QPOs is distinct, manifesting as a region of high bicoherence where both $f_{1}$ and $f_{2}$ are equal to $f_{\rm QPO}$, but without evident coupling between QPOs and the broadband noise. The systematic analysis of multiple sources using this method indicates a probable inclination-dependent alteration in the nonlinear properties of type-C QPOs. This phenomenon appears to be closely associated with the varying states observed during the evolution of outbursts in low-mass X-ray binaries \cite{arur2019non,arur2020likely}. \cite{arur2022using} applied this method to investigate the radio-X-ray correlations in GRS 1915+105. The study reveals that the bicoherence pattern correlates with the QPO frequency, hardness ratio, and radio properties. Moreover, phase coupling varies across different radio conditions, including radio-quiet, plateau, and steep conditions. Significant differences in phase lag behavior among observations with QPO frequencies above 2 Hz suggest a change in the underlying physical mechanism. 
Recently, \cite{ding2023nonlinear}  conducted a study using bicoherence analysis on MAXI J1820+070 and MAXI J1535-571. 
\cite{zhu2024bicoherence} also applied this method to study QPOs in MAXI J1535-571. They found that the bicoherence pattern of type C QPOs transitions from a 'web' pattern to a 'hypotenuse' pattern after the appearance of type B QPOs, suggesting that MAXI J1535-571 is a low-inclination source. Additionally, the intensity of bicoherence varies across different energy bands. 

Swift J1727.8$-$1613, initially identified as GRB 230824A, emerged as a new X-ray transient and was first detected by Swift's Burst Alert Telescope on August 24, 2023 \citep{Page2023, Kennea2023GCN}. Immediate monitoring using MAXI in the 2-20 keV energy range detected an X-ray flux exceeding 2 Crab \citep{negoro2023maxi}. Subsequent observations in optical \citep{castro2023optical}, radio \citep{miller2023vla}, and X-ray \citep{o2023nicer} wavelengths also suggest the source's candidacy as a black hole. Type-C QPOs have been observed in X-ray data \citep{palmer2023swift, draghis2023preliminary,sunyaev2023integral,katoch2023detection,bollemeijer2023nicer,mereminskiy2023hard}, and the Imaging X-ray Polarimetry Explorer (IXPE) has also conducted observations of this source \citep{dovciak2023ixpe,dovciak2023ixpeb}.

The orbital period of this source is approximately 7.6 hours, and it is located about $d=2.7\pm 0.3~\rm kpc$ away from us. Additionally, it exhibited complex signatures of optical inflows and outflows throughout its discovery outburst \citep{sanchez2024evidence}. 
\cite{veledina2023discovery}  the source with polarization degree (PD) $\sim$ 4.1\% ± 0.2\% and polarization angle (PA) $\sim$ 2.2$^\circ$ ± 1.3$^\circ$. Based on the polarization results and comparing X-ray flux with known sources, the inclination and distance of the source are predicted as i $\sim$ 30$^\circ$- 60$^\circ$ and 1.5 kpc, respectively. 
Additionally, the X-ray polarization angle aligns with observations in submillimeter wavelengths, indicating that the elongation of the corona is perpendicular to the direction of the jet \citep{ingram2023tracking}. \cite{wood2024swift} utilized the Very Long Baseline Array and the Long Baseline Array to image Swift J1727.8-1613. The images revealed a bright core and a large, two-sided, asymmetrical, resolved jet. Notably, the jet extends in the north-south direction, situated at a position angle of -0.60 $\pm$ 0.07$^\circ$ East of North. 
\cite{zhao2024first} conducted an analysis of QPOs in Swift J1727.8-1613 using IXPE data and the HHT method, revealing a notable modulation of the photon index corresponding to the QPO phase. However, inconsistencies arose as the PD and PA exhibited no such modulation, contradicting the expected results from the Lense-Thirring precession model. \cite{svoboda2024dramatic} reported that the X-ray polarization degree drops significantly to less than 1\% in the soft state, highlighting that X-ray polarization is largely dependent on the accretion state, with higher polarization in the hard state when emission is dominated by the X-ray corona

Numerous studies have also been conducted on the spectral properties of Swift J1727.8-1613. 
\cite{peng2024nicer} utilized simultaneous observations from Insight-HXMT, NICER, and the Nuclear Spectroscopic Telescope Array (\textit{NuSTAR}) to infer a spin parameter of approximately 0.98 and an orbital inclination of around 40 degrees for this source. 
Based on observations from \textit{Insight-HXMT}, the temporal and spectral analysis indicated that Swift J1727.8$-$1613 is a high-inclination, high-spin source \citep{yu2024timing, chatterjee2024insight,zhu2024energy}.

In this paper, we present the bicoherence analysis of type C QPOs in Swift J1727.8-1613 based on Insight-HXMT observations. Section \ref{obs} describes the observations and data reduction methods. Section \ref{result} provides the results of bicoherence. Sections ~\ref{DISCUSSION} and \ref{conclusion} present the discussions and conclusions, respectively.

\section{OBSERVATIONS AND DATA ANALYSIS}
\label{obs}
\subsection{Observations and Data Reduction}
After the outburst of Swift J1727.8$-$1613, Insight-HXMT conducted long-term observations of the source. Insight-HXMT is equipped with three types of detectors, each covering different energy ranges: the High Energy (HE) detectors operate between 20.0 and 250.0 keV with a temporal resolution of 4 $\rm \mu s$ \citep{liu2020High}, the Medium Energy (ME) detectors function in the 5.0 to 30.0 keV range with a temporal resolution of 240 $\rm \mu s$ \citep{cao2020medium}, and the Low Energy (LE) detectors cover 1.0 to 15.0 keV with a temporal resolution of 1 $\rm ms$ \citep{chen2020low}. The effective areas of these detectors are 5100 $\rm cm^2$, 952 $\rm cm^2$, and 384 $\rm cm^2$, respectively.

The data extraction and analysis were performed using Version 2.04 of the \textit{Insight-}HXMT Data Analysis Software (HXMTDAS)\footnote{\url{http://hxmten.ihep.ac.cn/software.jhtml}}. Good time intervals were determined by specific criteria: a pointing offset angle less than 0.04$^\circ$, an Earth elevation angle greater than 10$^\circ$, a geomagnetic cutoff rigidity above 8$^\circ$, and the exclusion of data within 300 seconds of passing through the South Atlantic Anomaly (SAA). Light curves were generated using the $helcgen$, $melcgen$, and $lelcgen$ tools in HXMTDAS. Background estimation was conducted with HEBKGMAP, MEBKGMAP, and LEBKGMAP, and further refined using $lcmath$ to subtract the estimated background noise. The analysis concentrated on energy ranges of 1--10 keV (LE), 10--20 keV (ME), and 20--100 keV (HE). 

\subsection{Timing analysis methods}
For each observation, we computed the PDS using the \textit{powspec} tool, employing a time interval of 64 seconds and a corresponding time resolution of 0.0078125 seconds. Subsequently, we averaged the obtained PDS for each observation and re-binned the resulting PDS using a geometric factor of 1.05 in frequency space. Normalization of the PDS was done in units of $\rm rms^2/Hz$, with Poisson noise subtracted \citep{belloni1990atlas, miyamoto1991x}. The total fractional variability (RMS of the PDS) within the 0.01–32 Hz range of LE was estimated.

We utilized multiple Lorentz functions to model both broadband noise (BBN) and QPO profiles within the PDS, employing XSPEC v12.14 \footnote{\url{https://heasarc.gsfc.nasa.gov/xanadu/xspec/manual/manual.html}}. This allowed us to extract the fundamental parameters of the QPO. To determine the fractional RMS of the QPO, we utilized the formula:
\begin{equation}
\rm rms_{\rm QPO}=\sqrt{R}\times\frac{S+B}{S},
\end{equation}
where S denotes the source count rate, R represents the normalization of the QPO's Lorentzian component, and B signifies the background count rate \citep{bu2015correlations}.

The bispectrum is a method of higher-order Fourier spectral analysis, utilized for examining the nonlinear characteristics present within time series data. This method is particularly well-suited for analyzing the variability data we have obtained. 
The bispectrum of a time series divided into K intervals is:
\begin{equation}
B(k, l) = \frac{1}{K} \sum_{i=0}^{K-1} X_i(k)X_i(l)X^*_i(k + l), 
\end{equation}
where \(X_i(f)\) is the Fourier transforming of the \(i\)th segment of the time series at frequency \(f\), and \(X^*_i(f)\) is the complex conjugate of \(X_i(f)\).
The bispectrum is a complex number, with its real component probing the skewness of the underlying flux distribution and its imaginary component examining the asymmetry of the waveform \citep{maccarone2013biphase}. 

A related term is bicoherence, which is similar to the cross-coherence function. The bicoherence ranges from 0 to 1, where 0 signifies no nonlinear coupling between the phases of different Fourier frequencies, and 1 signifies complete coupling. 
Using the normalization proposed by \cite{kim1979digital}, the bicoherence is calculated as:
\begin{equation}
b^2 = \frac{\left| \sum X_i(k) X_i(l) X^*_i(k + l) \right|^2}{\sum |X_i(k) X_i(l)|^2 \sum |X_i(k + l)|^2} ,
\end{equation}
$b^2$ measures the fraction of power at the frequency $ k + l $ attributed to the coupling of the three frequencies, $k$, $l$ and $ k + l $ .
Similar to previous studies\citep{arur2020likely,zhu2024bicoherence}, following the computation of bicoherence values for various observations at different frequencies $f_1$ and $f_2$ (representing the frequencies $k$ and $l$ ), we utilized $f_1$ and $f_2$ as axes, with colors indicating different values of $\log b^2$ at various points. 

In our calculations, we selected 16 seconds as the length of each interval, corresponding to a frequency resolution of 0.0625 Hz. We fitted the QPO phenomena for all observations and chose observations with  QPO frequency greater than 1 Hz to ensure that the signal exceeded the resolution. Additionally, as pointed out by \cite{arur2020likely}, segments need to be long enough to get sufficient cycles for analysis. Therefore, we selected observations with an exposure time greater than 960 seconds (corresponding to about 60 intervals). Additionally, we used the Augmented Dickey-Fuller technique to check the selected observations to ensure they are stationary \citep{hamilton2020time}.
Ultimately, we selected 133 observations starting from observation ID P061433800307 for analysis. Additionally, We calculated the hardness ratio for each observation, where hardness is defined as the ratio of background-subtracted mean count rates between 4.0–12.0 keV and 2.0–4.0 keV.

\section{RESULT} \label{result}
\subsection{QPO phenomena and hardness ratio}
\begin{figure*}
    \includegraphics[width=\columnwidth]{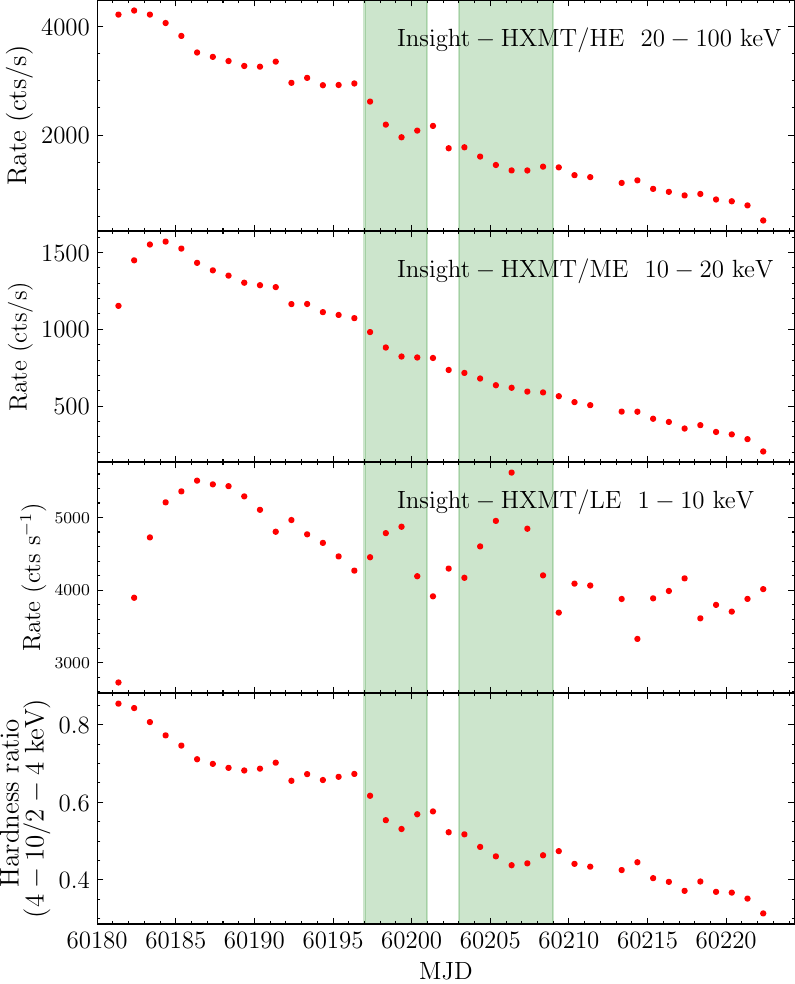}
    \includegraphics[width=\columnwidth]{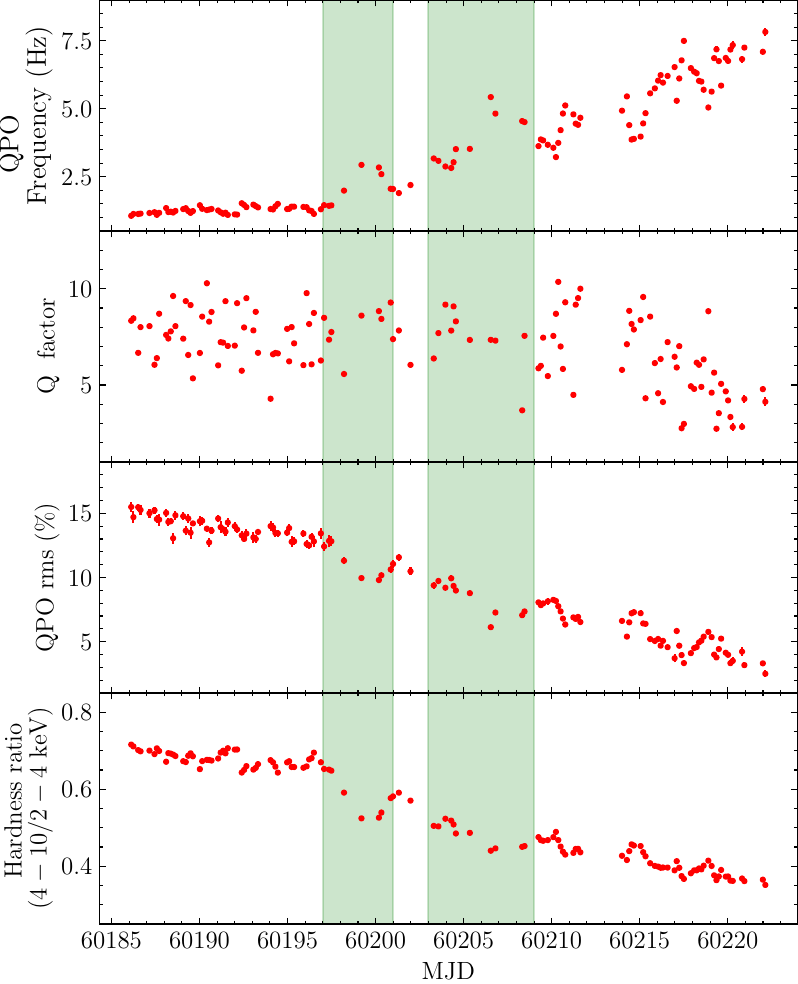}
    \caption{Left-hand panel:from top to bottom, \textit{Insight}-HXMT HE (20--100 keV) data, \textit{Insight}-HXMT ME (10--20 keV) data,
    \textit{Insight}-HXMT LE (1--10 keV) data,
    and hardness ratios derived from \textit{Insight}-HXMT data using  (4--10 keV)/(2--4 keV), each data point in this panel is the result of rebinning the observation data with a time step of one day. 
    Right-hand panel: from top to bottom, the type-C QPOs' fundamental frequency, Q factor, and fractional RMS from the LE energy band and the hardness ratio, are shown as functions of time for Swift J1727.8-1613. The first green region spans from 60197 to 60201, which we refer to as flare-1. The second green region extends from 60203 to 60209, which we refer to as flare-2.  The uncertainties in the data are generally about 2 orders of magnitude smaller than the data themselves; therefore, the errorbars in the figure are very short and difficult to distinguish.}
    \label{figure1}
\end{figure*}

In Figure~\ref{figure1}, the left panel displays the light curves and hardness ratios in different energy bands during the outburst process, with each data point representing the results rebinned to one day. On the right panel, we have presented the results of the selected QPO parameters from the LE energy band as they have evolved over time from $\sim$ 1 Hz to $\sim$ 8 Hz, with each data point representing the results of one observation. 

In the light curve of the HE energy band, it is evident that the count rate has remained consistently high since the beginning of the observation, gradually declining afterward. However, during the two flare periods, a distinct pattern emerges, characterized by an initial decrease followed by an upward trend.
In both the ME and LE energy bands, similar patterns of change are observed in the count rate. Initially, there is an increase followed by a gradual decline. However, during the descent in the LE energy band, two flare events are observed (marked by green regions). 
The hardness ratio indicates a gradual softening of the source. During the two flare periods, there is a trend of initial softening followed by a slight hardening.

The frequency of the QPO gradually increases over time, undergoing a rapid rise and subsequent rapid decline during the flare period. 
The Q factor shows no significant variation over time before MJD 60212, consistently remaining between 5 and 10. However, after MJD 60212, it exhibits a general decreasing trend over time. The value of the QPO rms gradually decreases over time.

\begin{figure}
    \includegraphics[width=\columnwidth]{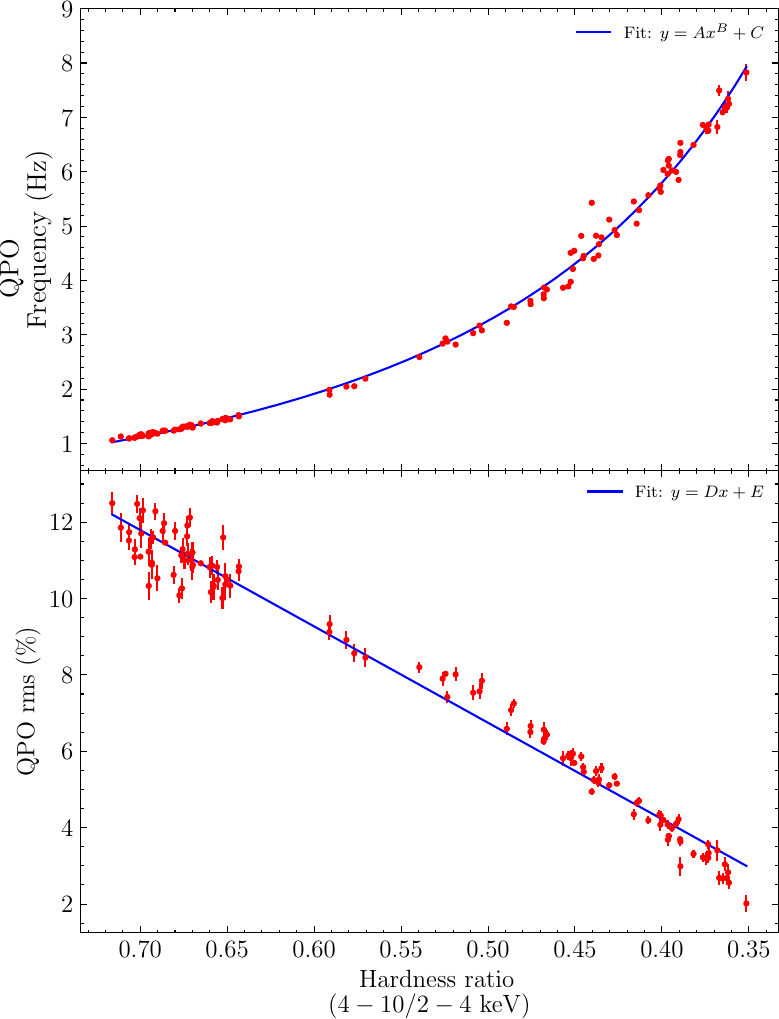}
    
    \caption{Top panel: the relationship between QPO frequency and hardness ratio. The blue curve in the figure represents our fitting result using the equation 
$y=\mathrm{A}x^{\mathrm{B}}+\mathrm{C}$.
    Bottom panel: the relationship between QPO RMS and hardness ratio, with the blue line representing the best fitting linear function $y=\mathrm{D}x+\mathrm{E}$. }
    \label{figure2}
\end{figure}

In Figure~\ref{figure2}, we plotted the QPO frequency as a function of hardness ratio in the top panel and the QPO RMS as a function of hardness ratio in the bottom panel for these observations we selected of Swift J1727.8-1613. 
The QPO frequency generally increases as the hardness ratio decreases and the QPO RMS 
decreases as the hardness ratio decreases. We used the equation 
$y=\mathrm{A}x^{\mathrm{B}}+\mathrm{C}$ to fit the relationship between QPO frequency and hardness ratio. The obtained parameter values are 
$\mathrm{A}=0.95\pm 0.09$, $\mathrm{B}=-2.13\pm 0.08$ and  $\mathrm{C}=-0.92\pm 0.15$. 
Given the approximately linear relationship between QPO RMS and hardness ratio, we utilized the linear function $y=\mathrm{D}x+\mathrm{E}$ for fitting. The obtained fitting parameters are 
$\mathrm{D}=25.23\pm 0.33$ and $\mathrm{E}= -5.87\pm 0.19$. 

\subsection{Bicoherence patterns}
We classified the different phenomenological patterns observed in the bicoherence according to the conventions established by \cite{maccarone2011coupling}. The 'hypotenuse' pattern appears when high bicoherence is observed in the diagonal region, where the sum of the two frequency components equals the QPO frequency. The 'cross' pattern is characterized by high bicoherence observed for frequency pairs, where one frequency corresponds to the QPO and the other frequency varies freely. The ‘web’ pattern represents a hybrid class, exhibiting both the diagonal characteristic of the ‘hypotenuse’ and the vertical and horizontal streaks characteristic of the ‘cross’ pattern. 

\subsubsection{Bicoherence patterns in LE}

\begin{figure*}
    \includegraphics[scale=0.455]{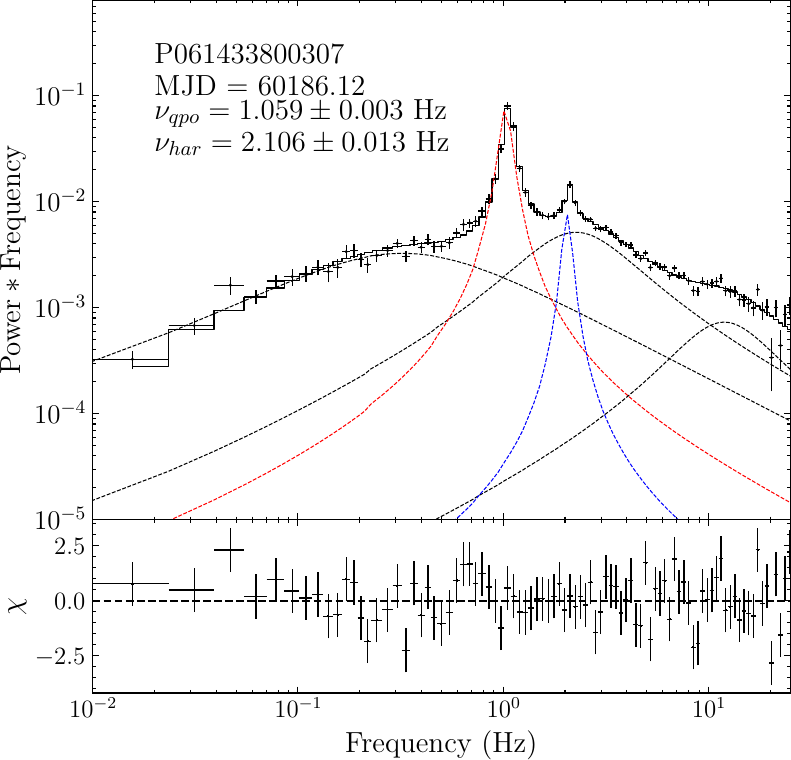}
    \includegraphics[scale=0.455]{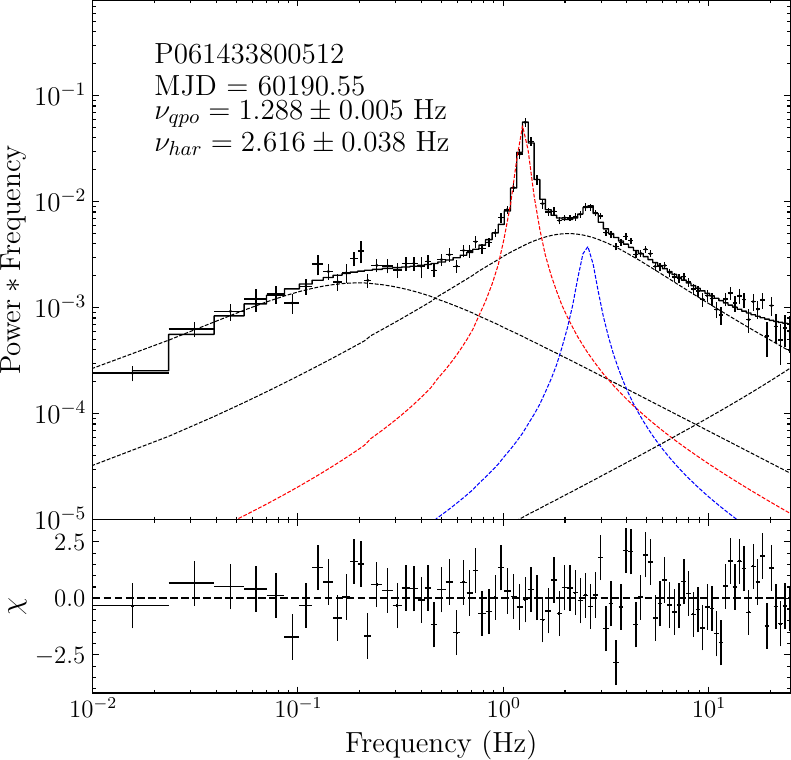}
    \includegraphics[scale=0.455]{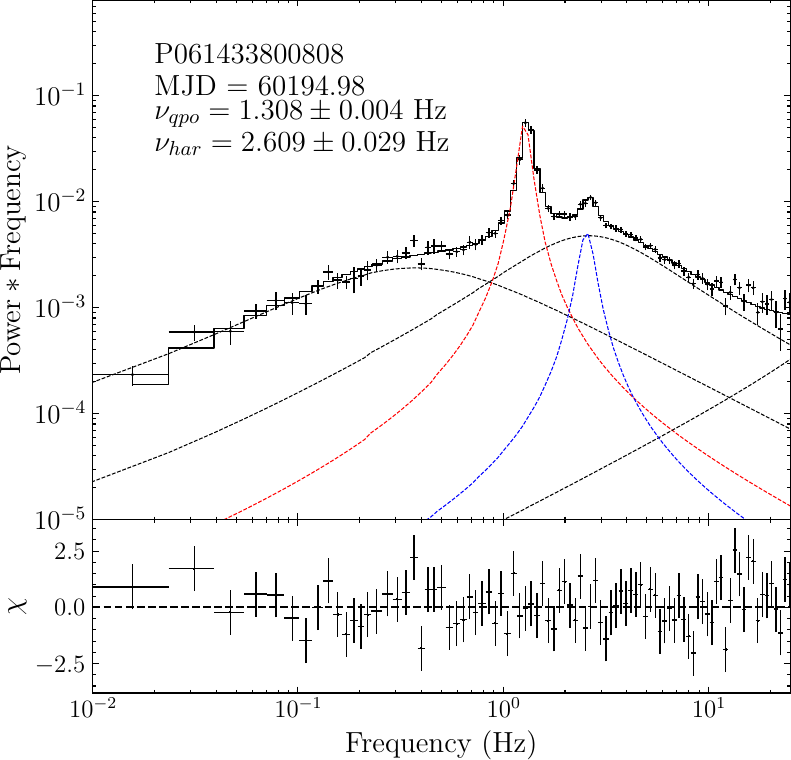}\\
    \includegraphics[scale=0.455]{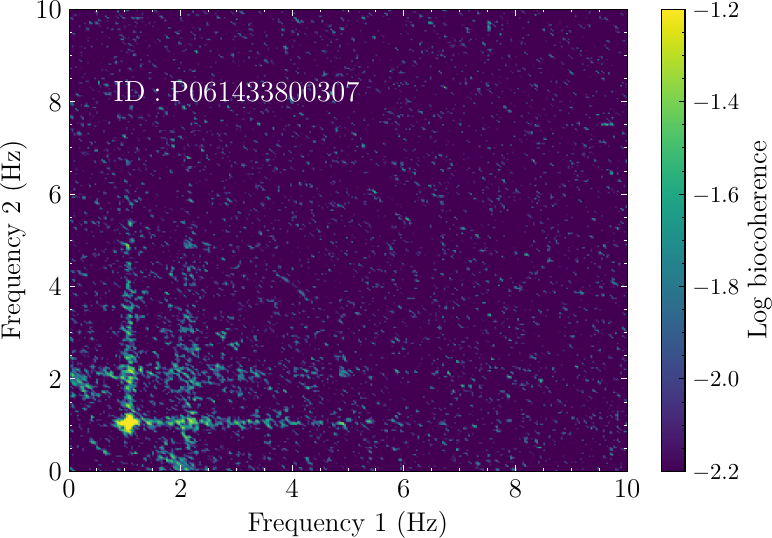}
    \includegraphics[scale=0.455]{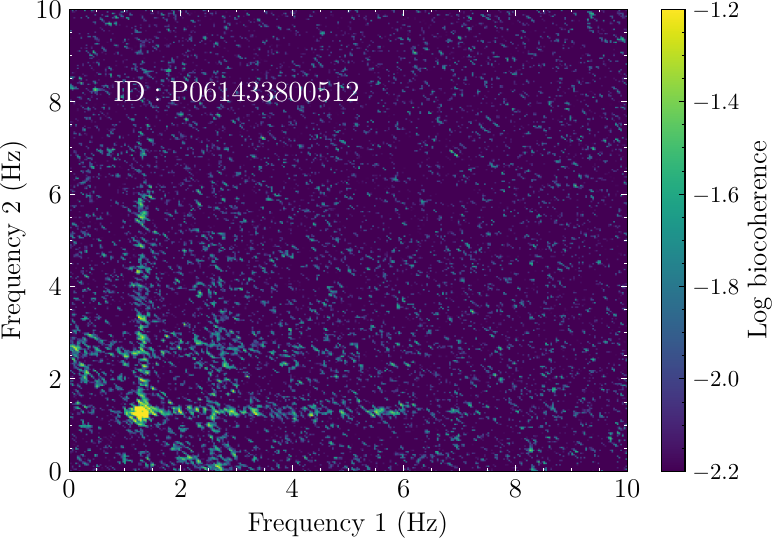}
 \includegraphics[scale=0.455]{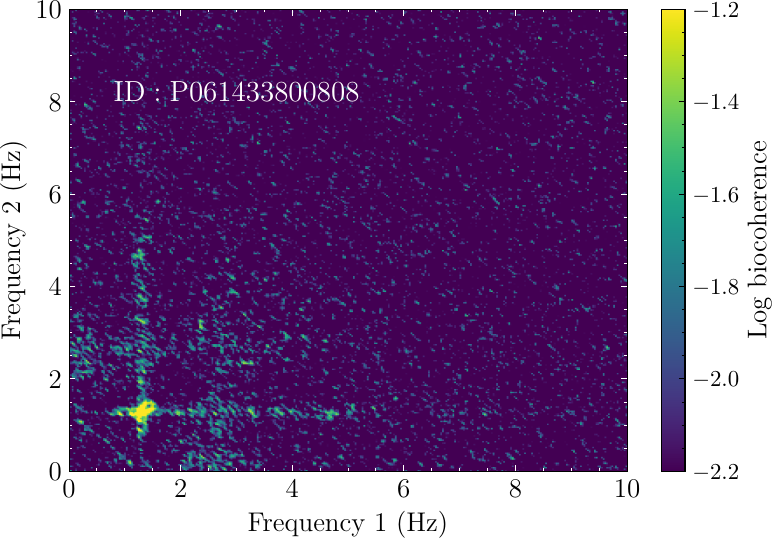}\\
      \includegraphics[scale=0.455]{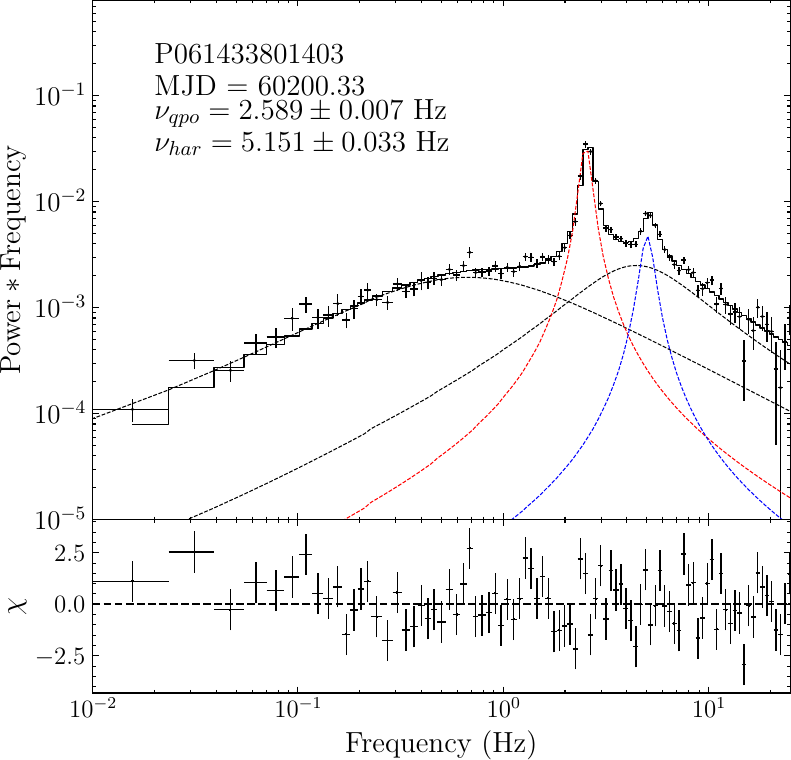}
    \includegraphics[scale=0.455]{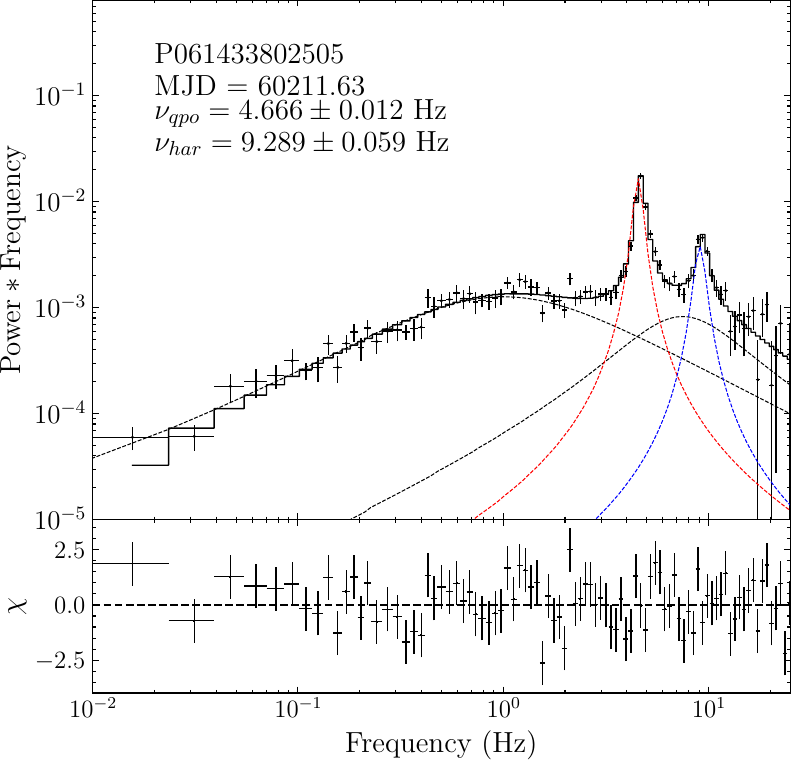}
    \includegraphics[scale=0.455]{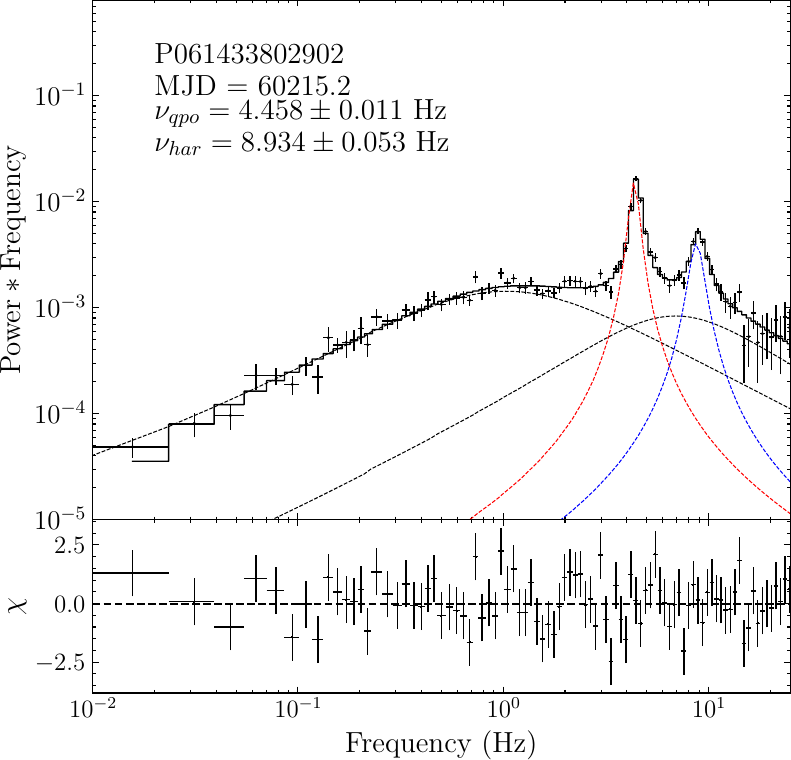}\\
    \includegraphics[scale=0.455]{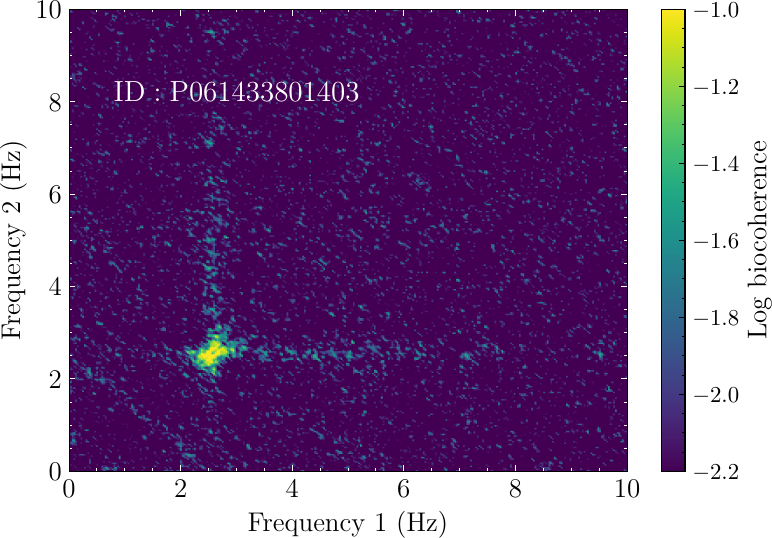}
    \includegraphics[scale=0.455]{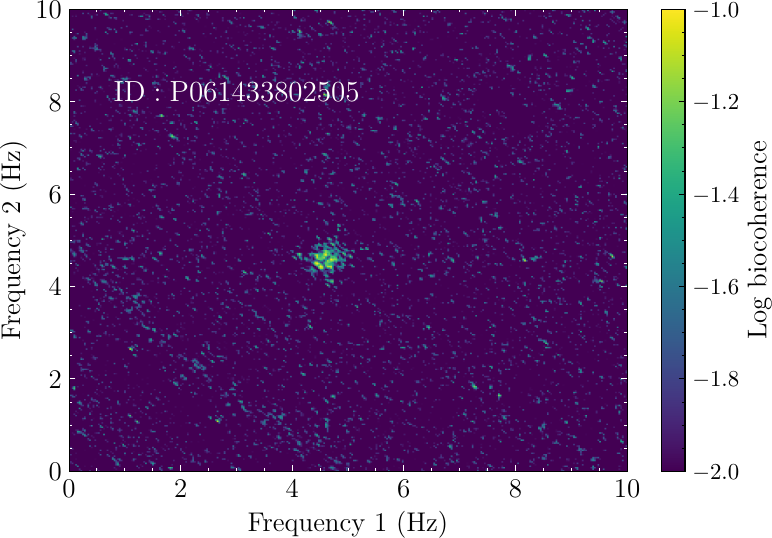}
 \includegraphics[scale=0.455]{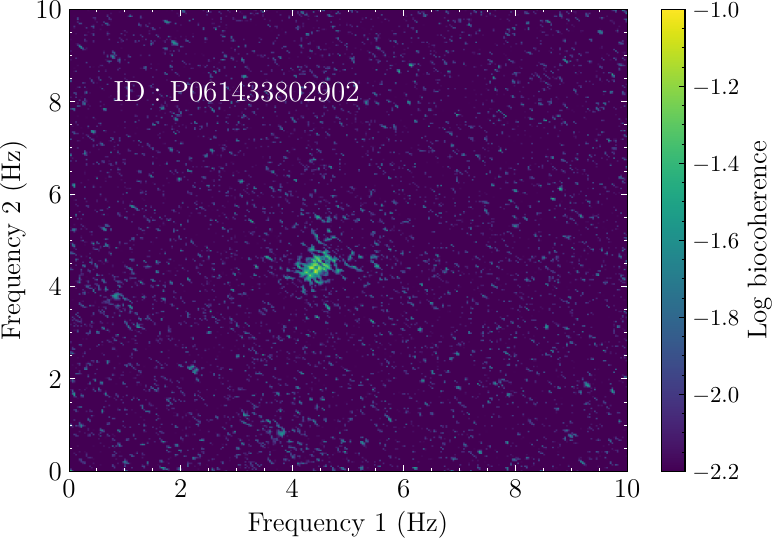}\\

    \caption{The figures show representative bicoherence results in LE. The first and third rows display the observed PDS, where the red lines represent the QPO fundamental component and the blue lines denote the harmonics component. The second and fourth rows show the bicoherence results. The observation ID, time, and the fitted frequency values for each observation are labeled in each subplot. 
}
    \label{figure3}
\end{figure*}

During the outburst of Swift J1727.8-1613, multiple forms of bicoherence were observed. We presented the representative results in Figure~\ref{figure3}. The first and third rows have displayed the PDS, while the second and fourth rows have shown the bicoherence patterns.

During the initial phase, before entering the flare state, the bicoherence pattern exhibited characteristics of the web pattern, high bicoherence was observed at the frequencies of QPOs and harmonics. Relatively high bicoherence is also found where two frequency components add up to the frequency of the QPO, although it is not prominent in some observations (e.g., Observation ID: P061433800512). Additionally, it is worth noting that relatively high bicoherence also occurs when 
$f_{1}+f_{2}=f_{\rm har}$. However, this structure only appears at the endpoints of the diagonal, disappearing around $f_{1}=f_{2}=f_{\rm QPO}$. Moreover, it gradually weakens over time. 

After entering the flare state (MJD=60197), it is evident that the high bicoherence at 
$f_{1}=f_{2}=f_{\rm har}$ along the horizontal and vertical directions disappears. Additionally, the diagonal structure at $f_{1}+f_{2}=f_{\rm QPO}$ becomes more pronounced. As time progresses, the bicoherence at $f_{1}=f_{2}=f_{\rm QPO}$ gradually weakens. The diagonal feature persists but also diminishes gradually. In the final subplot of Figure 3, after the displayed observation, the bicoherence becomes very weak, and no discernible pattern is evident. 

During the outburst of Swift J1727.8-1613, our results presented in Figure~\ref{figure3} clearly demonstrate the transition of the bicoherence pattern from the 'web' pattern to the 'hypotenuse' pattern in the LE energy band. 

\subsubsection{Bicoherence patterns in ME}

\begin{figure*}
    \includegraphics[scale=0.45]{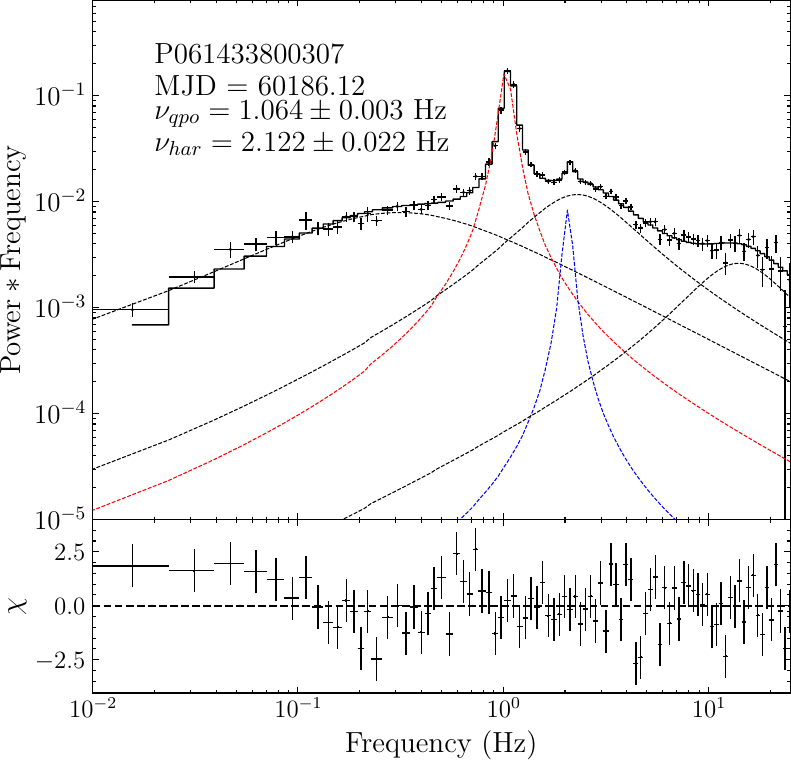}
    \includegraphics[scale=0.45]{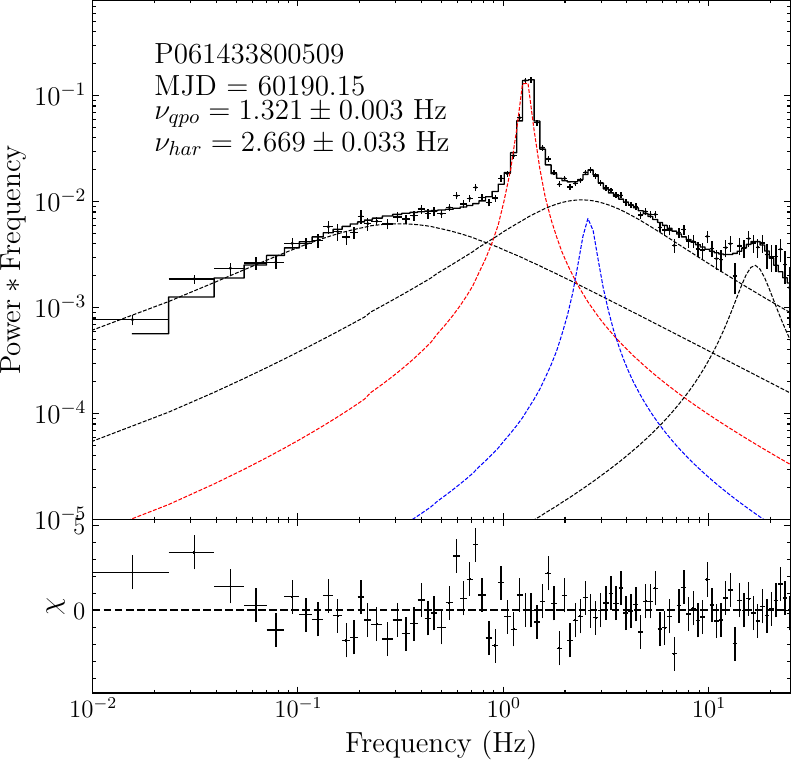}
    \includegraphics[scale=0.45]{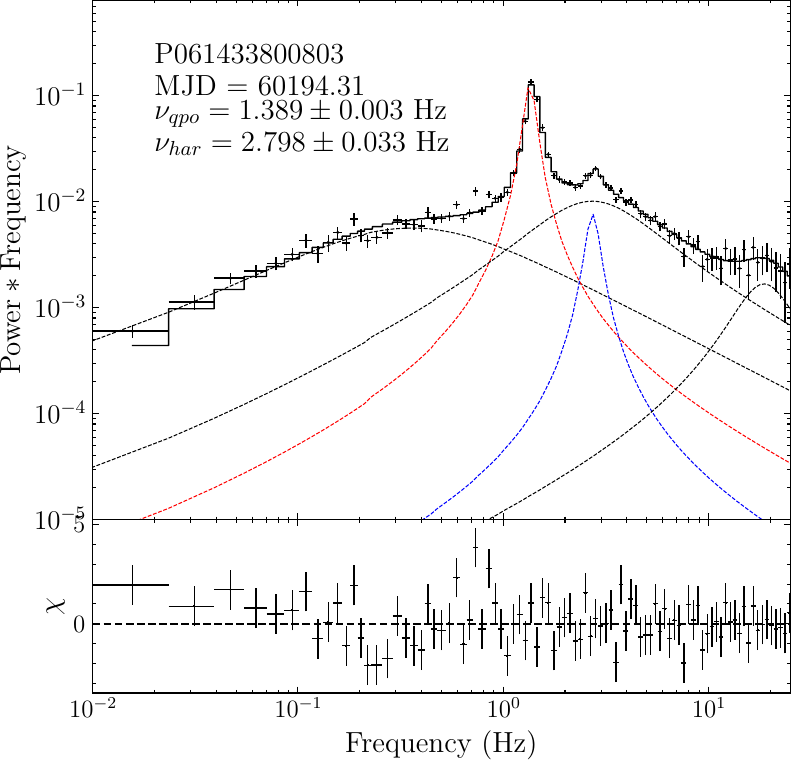}\\
    \includegraphics[scale=0.45]{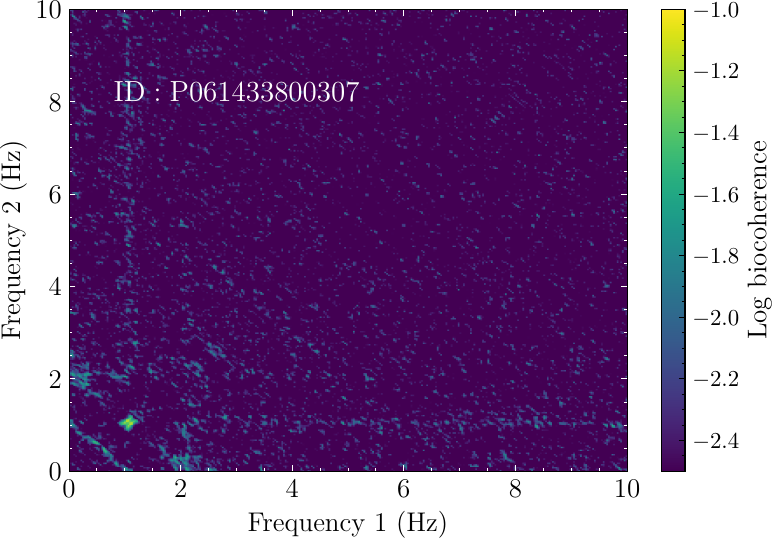}
    \includegraphics[scale=0.45]{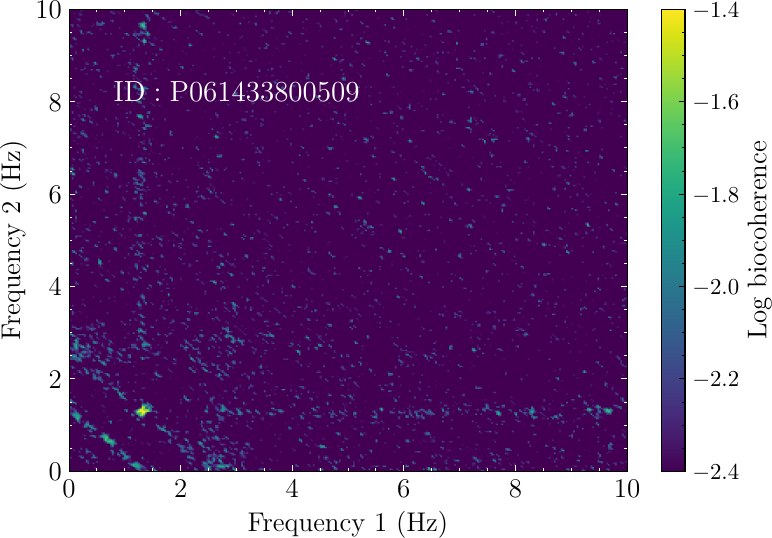}
 \includegraphics[scale=0.45]{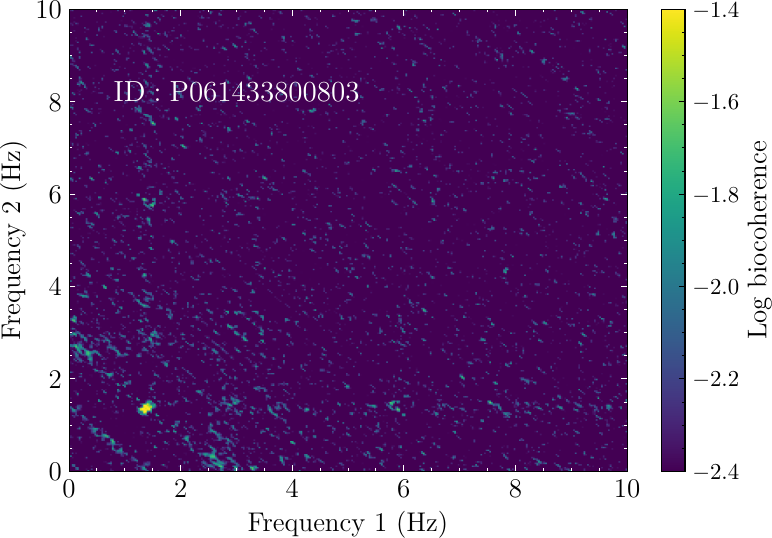}\\
      \includegraphics[scale=0.45]{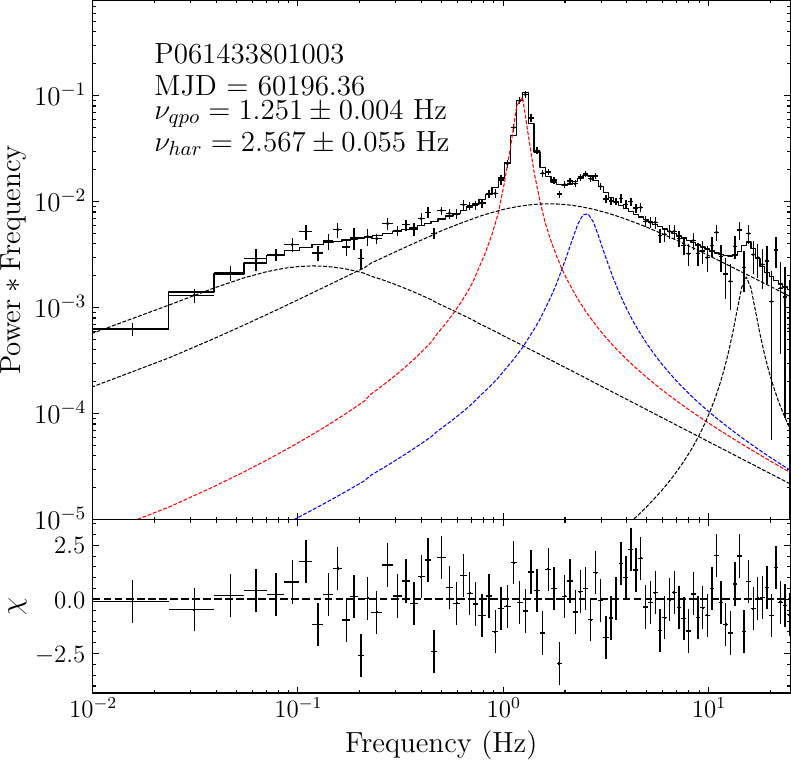}
    \includegraphics[scale=0.45]{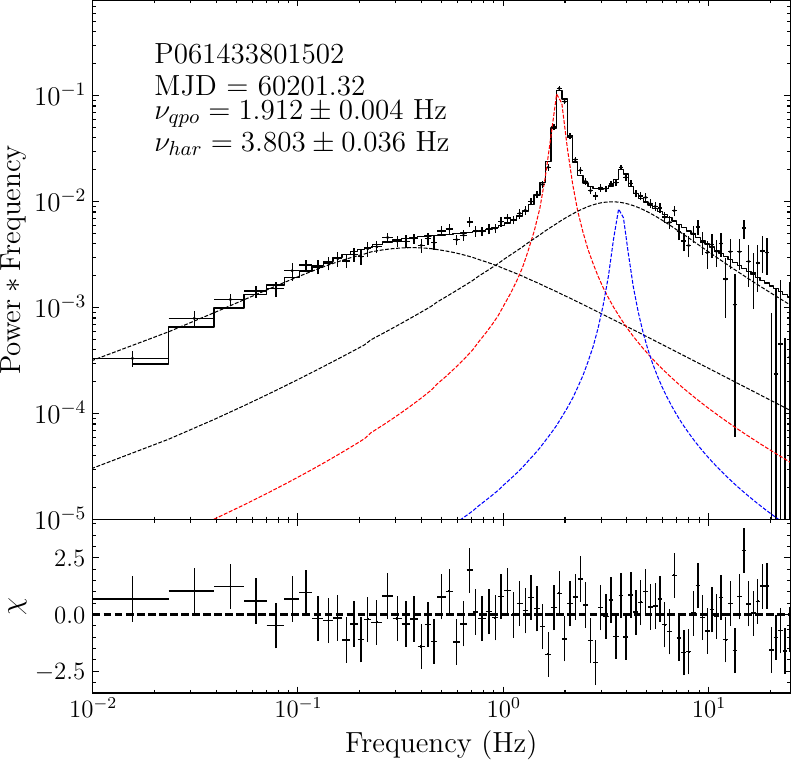}
    \includegraphics[scale=0.45]{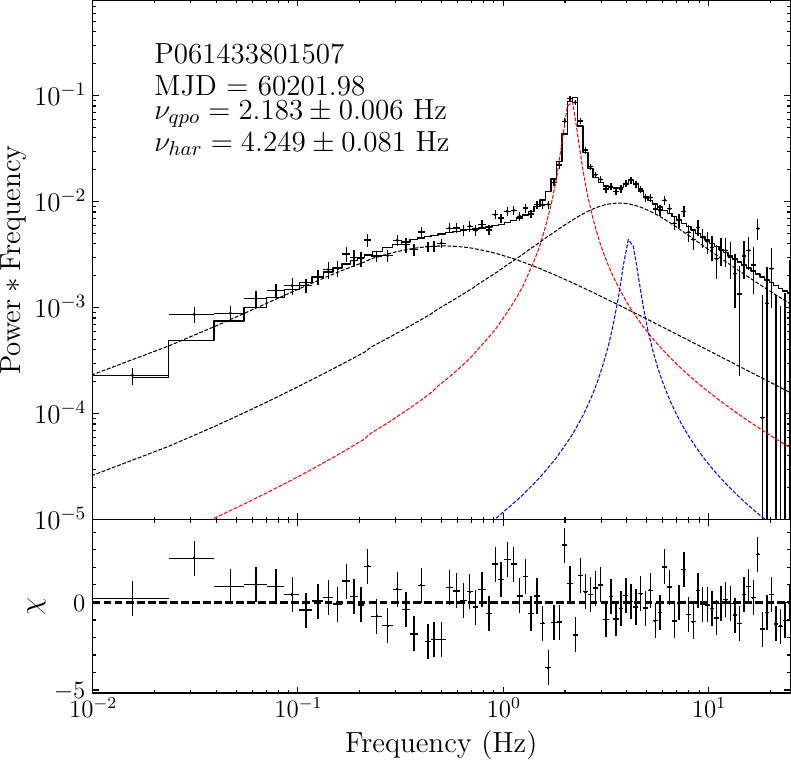}\\
    \includegraphics[scale=0.45]{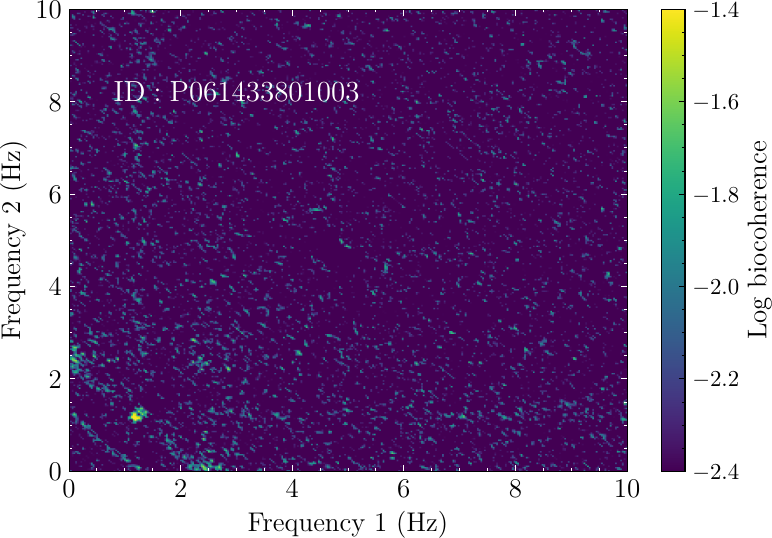}
    \includegraphics[scale=0.45]{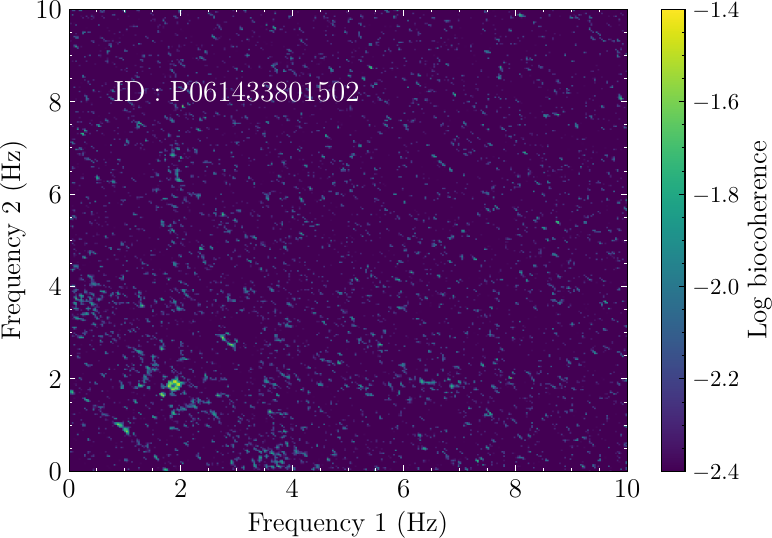}
 \includegraphics[scale=0.45]{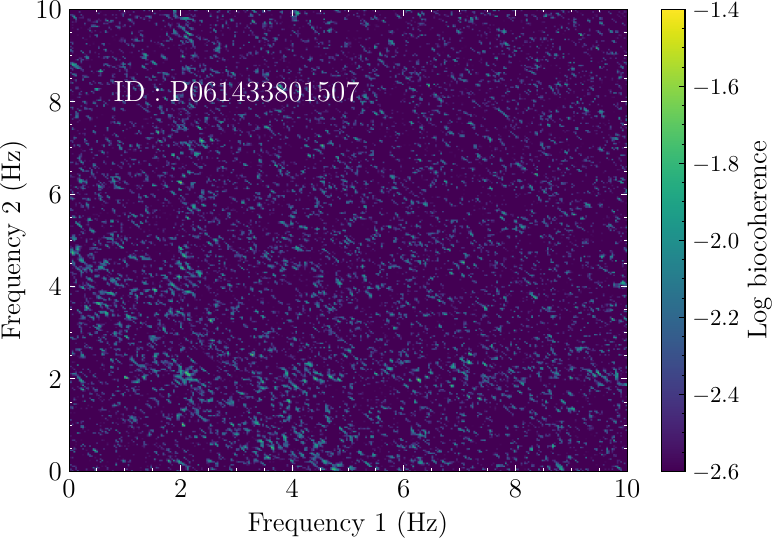}\\

    \caption{The figures show representative bicoherence results in ME. The first and third rows display the observed PDS. The second and fourth rows show the bicoherence results. The blue and red lines represent the QPO components and harmonic components, respectively. The observation ID, time, and the fitted frequency values for each observation are labeled in each subplot. 
}
    \label{figure4}
\end{figure*}

In Figure~\ref{figure4},  we have presented exemplary results obtained in the ME energy band, with markings consistent with those in Figure~\ref{figure3}. In observations with discernible patterns, we have found that the bicoherence pattern remained nearly unchanged. Moreover, this bicoherence pattern differs from the three previously identified bicoherence patterns. 

At $f_{1}+f_{2}=f_{\rm QPO}$, there is a distinct high bicoherence. Additionally, the pattern where $f_{1}=f_{2}=f_{\rm QPO}$   persisted until the frequency exceeded 2 Hz, after which it disappeared. Compared to the bicoherence patterns in the LE band, the strength of the bicoherence in the vertical and horizontal streaks weakens, while it is higher in the diagonal structure, especially exhibiting a very distinct pattern at 
$f_1 + f_2 = f_{\rm har}$.  This diagonal structure runs parallel to the diagonal formed by $f_1 + f_2 = f_{\rm QPO}$. 
To distinguish this pattern, we refer to it as the 'parallel' pattern.

During the outburst of this source, the bicoherence in the ME band consistently maintained the "parallel" pattern in observations with discernible patterns. Over time, the bicoherence pattern does not show a change from one to another. Notably, the bicoherence at 
$f_1=f_2=f_{\rm QPO}$ disappeared earlier than the two diagonal lines in the bicoherence pattern. Subsequently, the disappearance of 
$f_{1}+f_{2}=f_{\rm har}$ preceded that of $f_1+f_2=f_{\rm QPO}$. 
In the observations after the one with ID P061433801507, no discernible pattern is observed in the bicoherence. 

\subsubsection{Bicoherence patterns in HE}
\begin{figure*}
    \includegraphics[scale=0.45]{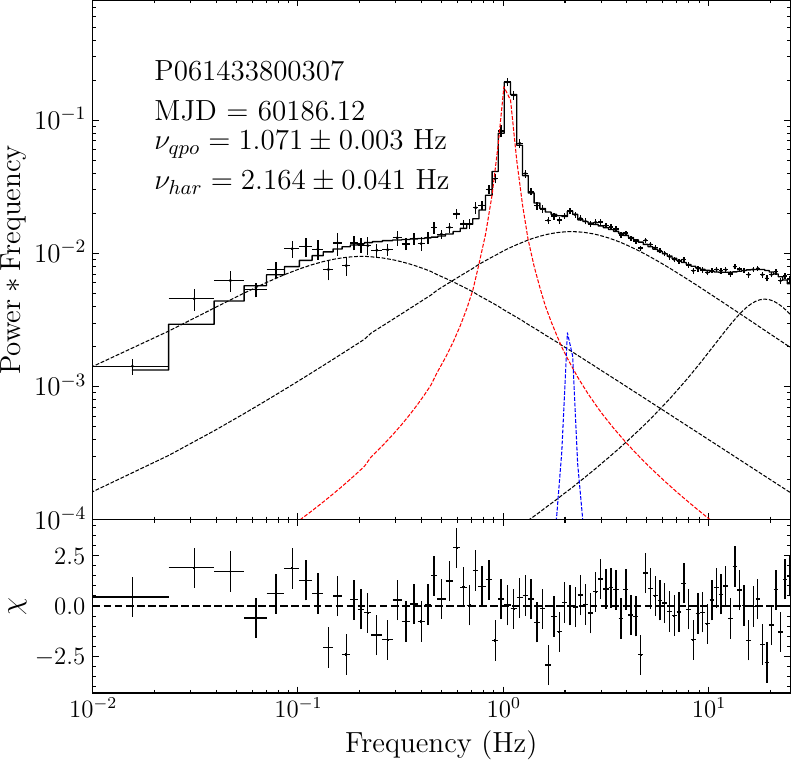}
    \includegraphics[scale=0.45]{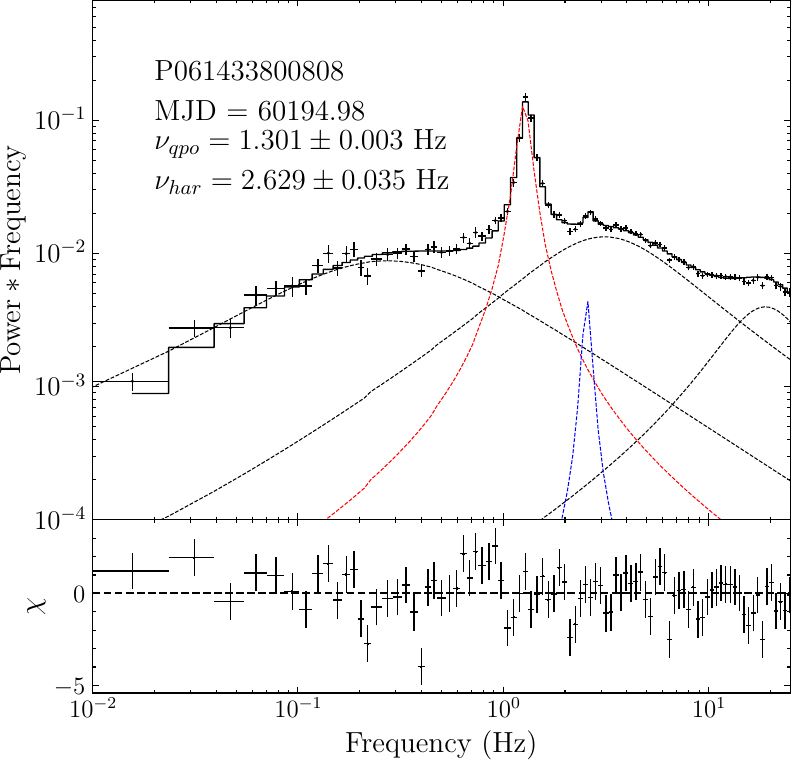}
    \includegraphics[scale=0.45]{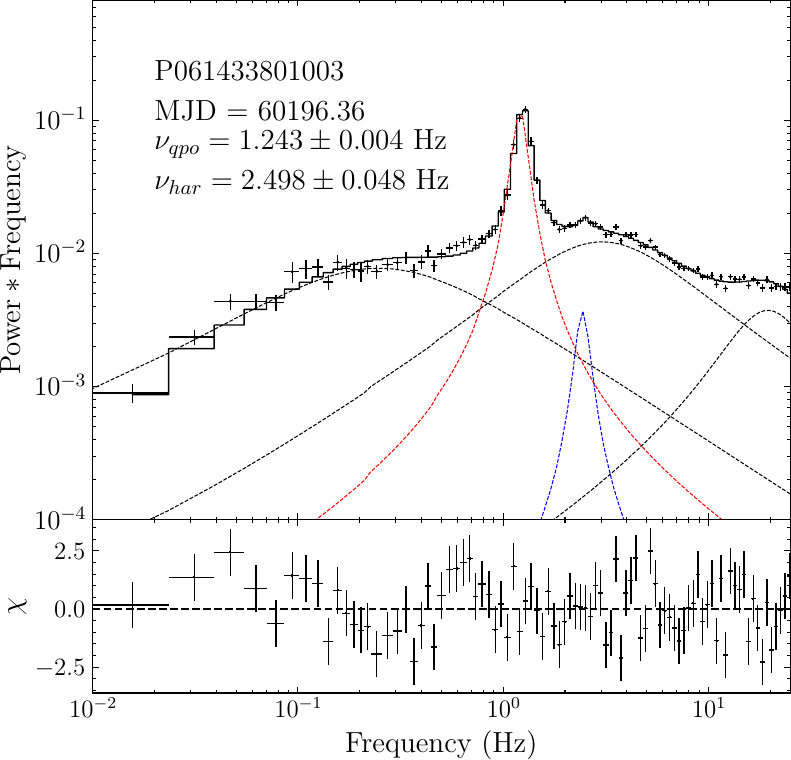}\\
    \includegraphics[scale=0.45]{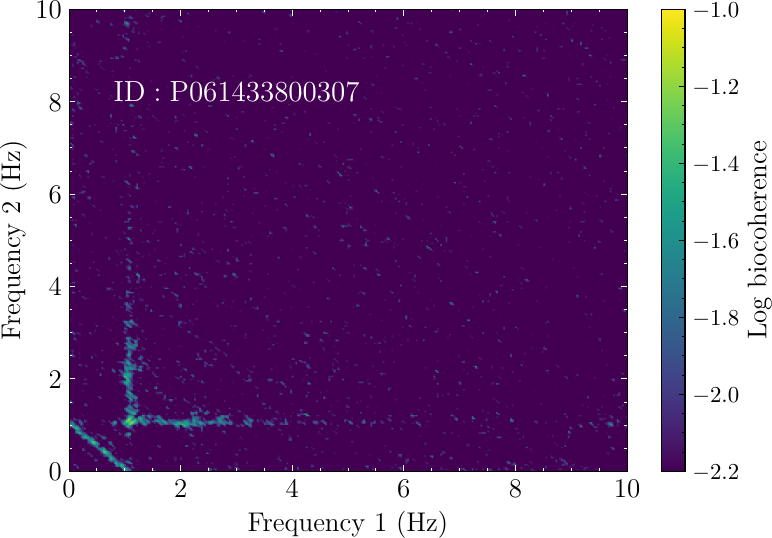}
    \includegraphics[scale=0.45]{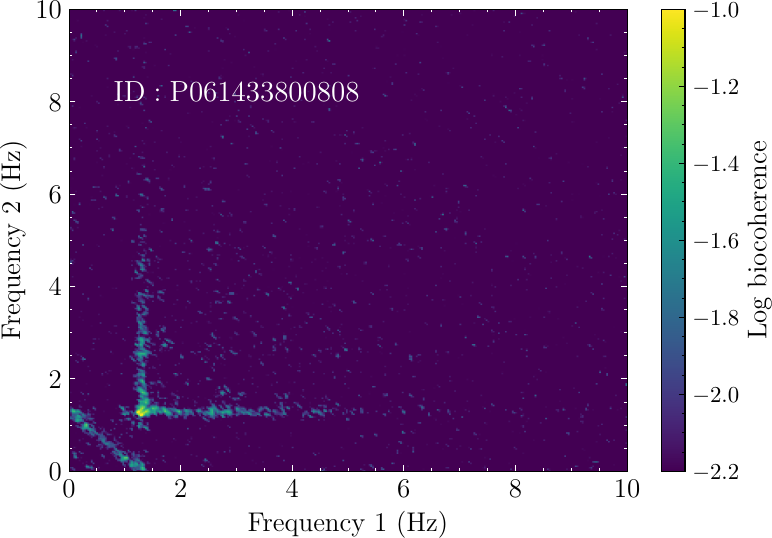}
 \includegraphics[scale=0.45]{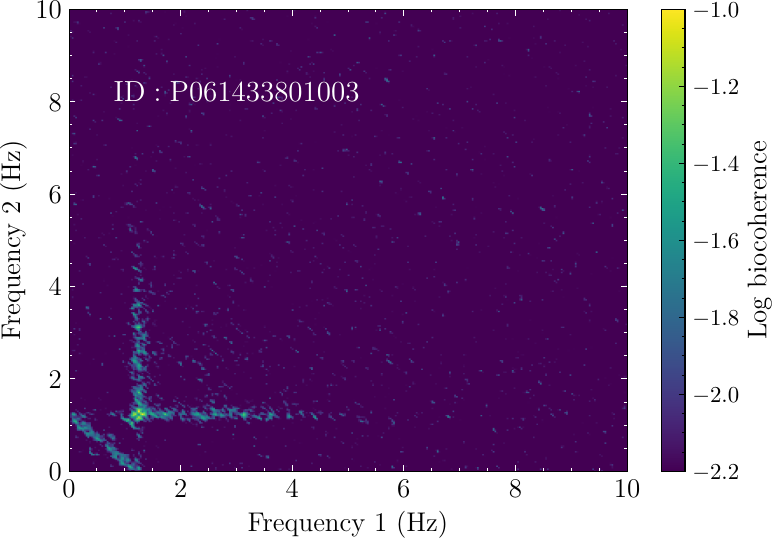}\\
      \includegraphics[scale=0.45]{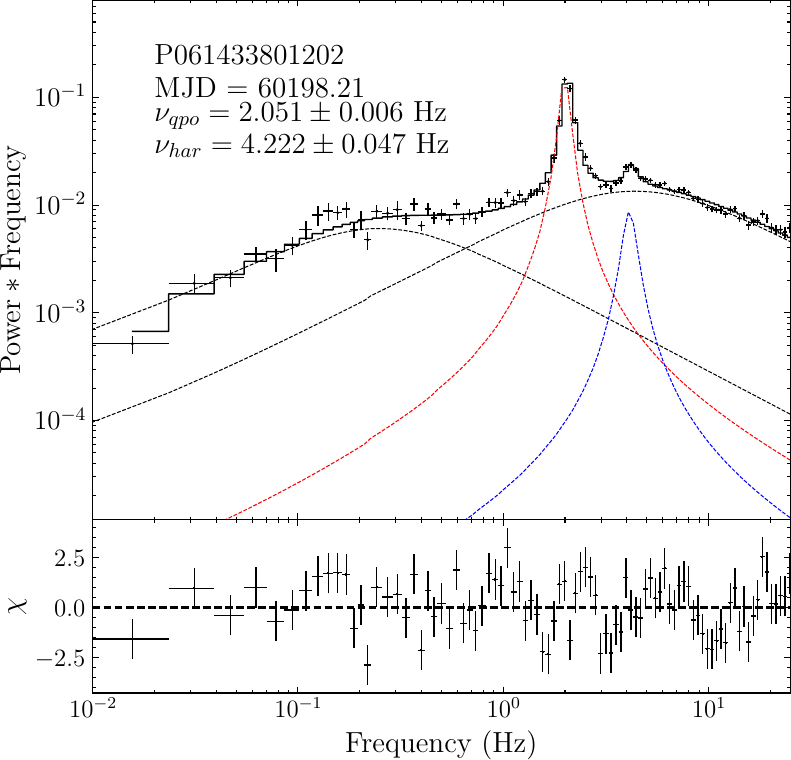}
    \includegraphics[scale=0.45]{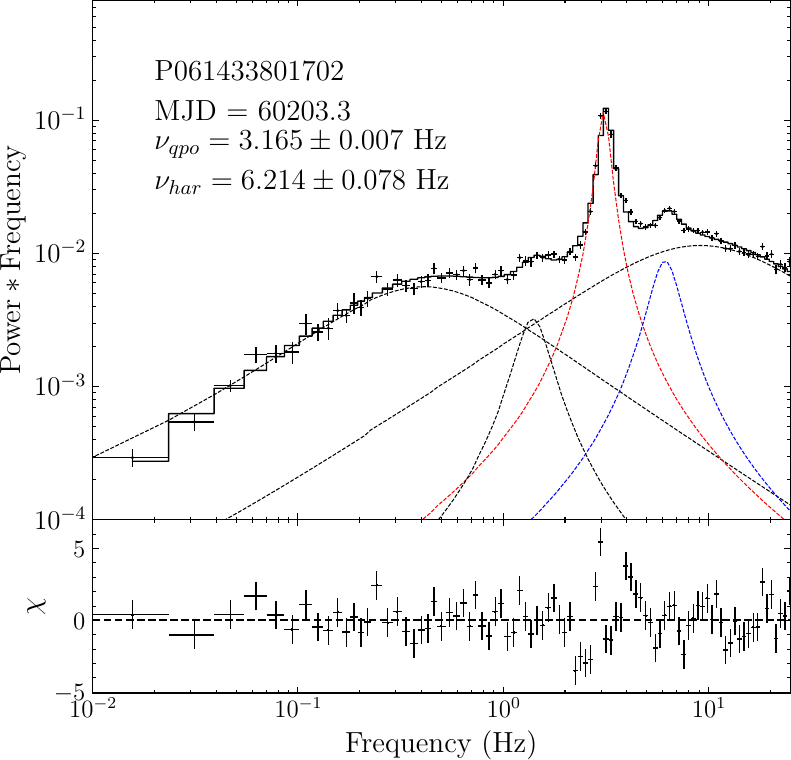}
    \includegraphics[scale=0.45]{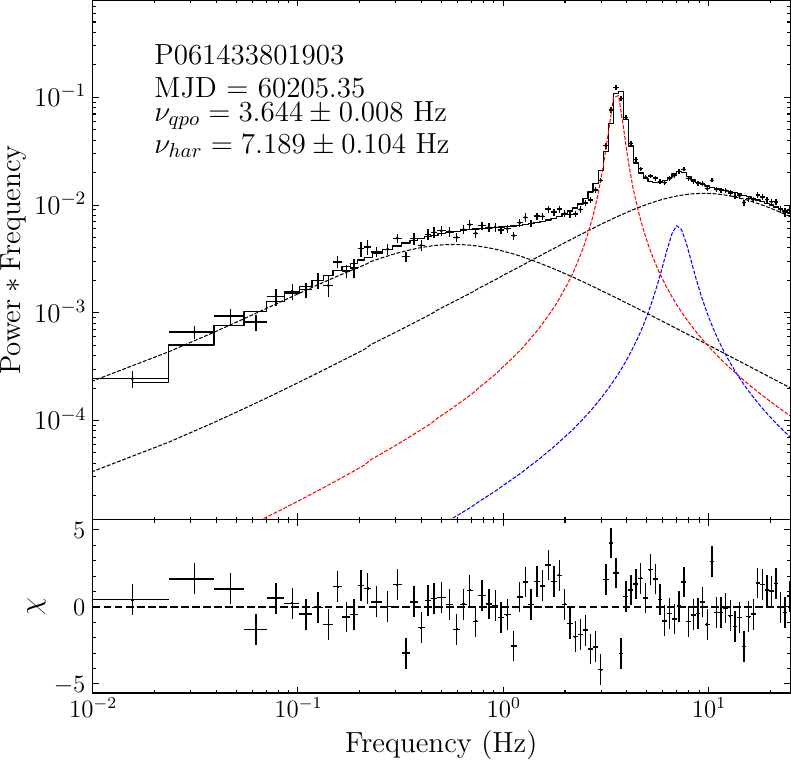}\\
    \includegraphics[scale=0.45]{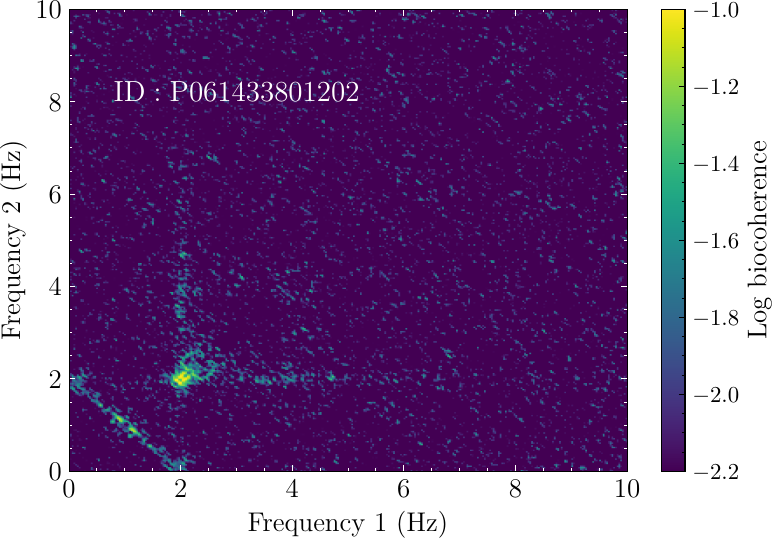}
    \includegraphics[scale=0.45]{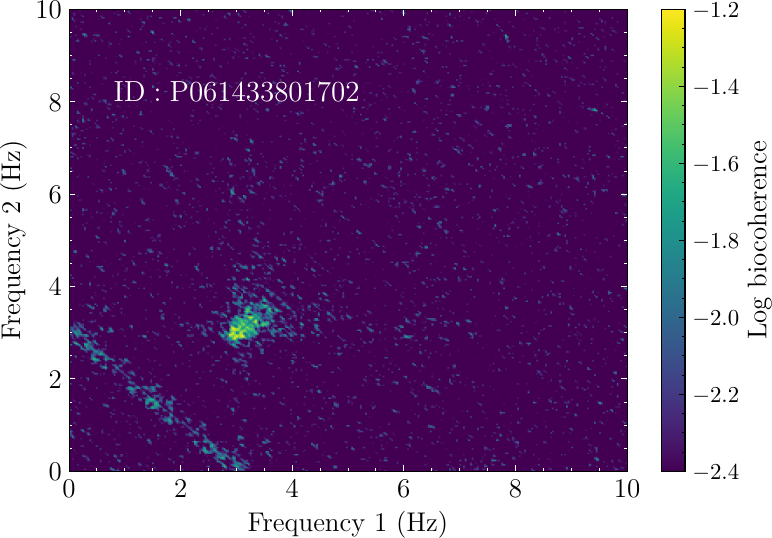}
 \includegraphics[scale=0.45]{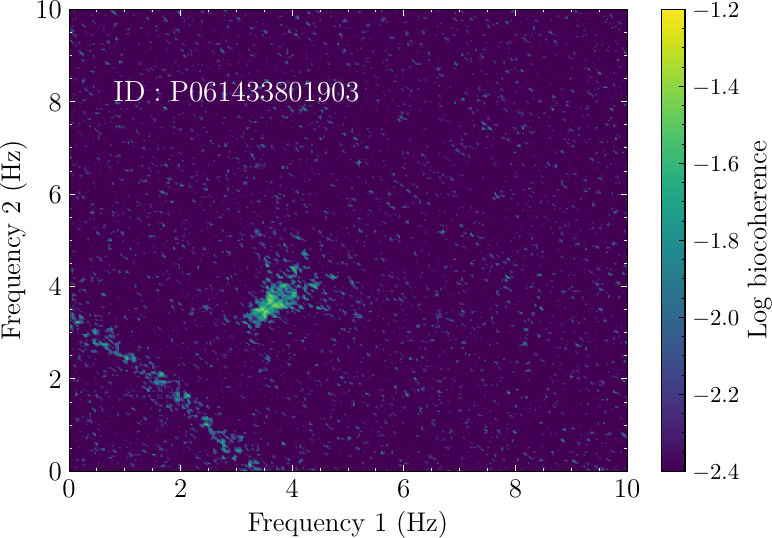}\\

    \caption{The figures show representative results in HE. The first and third rows display the observed PDS. The second and fourth rows show the bicoherence results. The blue and red lines represent the QPO components and harmonic components, respectively. The observation ID, time, and the fitted frequency values for each observation are labeled in each subplot. 
}
    \label{figure5}
\end{figure*}

Consistent with the results fitted by \cite{zhu2024energy}, we also found that the intensity of harmonics in the HE band becomes very weak, and the harmonic phenomenon disappears in some observations. We selected observations with obvious harmonic phenomena and presented the results along with the bicoherence results in Figure~\ref{figure5}. 

As can be observed, the evolution of the bicoherence pattern in the HE band differs from that in the LE and ME bands. It exhibits a 'web' pattern from the initial stage and persists for a very long time. This pattern remains unchanged upon entering the 'flare-1'.  However, after entering flare-2, the bicoherence pattern transformed from the 'web' pattern to the 'hypotenuse' pattern. 

Throughout the evolution of the bicoherence pattern during the emergence of the type C QPOs, both the HE and LE energy bands transform from the 'web' pattern to the 'hypotenuse' pattern. However, at $f_{1}=f_{2}=f_{har}$, the bicoherence is not clearly observed due to the weakening of the harmonic strength.

\section{DISCUSSION} \label{DISCUSSION}
\subsection{Inclination of Swift J1727.8-1613 }
We conducted a comprehensive investigation of the bicoherence phenomenon of type-C QPOs during the outburst of Swift J1727.8-1613. Our findings revealed a dynamic evolution in the bicoherence pattern of type-C QPOs. Similar pattern variations have been observed in previous studies involving other sources using similar methods. Notably, these pattern changes exhibit a correlation with the inclination angle. 

\cite{arur2019non} reported that in their study of GX 339-4, the bicoherence pattern changes gradually from web to hypotenuse as the source transitions from a hard intermediate state to a soft intermediate state. Subsequently, \cite{arur2020likely}  conducted a follow-up study to further investigate the relationship between the evolution of bicoherence and the inclination angle. 
They observed that the non-linear properties of type C QPOs vary with inclination as the source transitions from a LHS to a SIMS. For high inclination sources, the pattern changes from a web to a cross, while for low inclination sources, it changes from a web to a hypotenuse. 
\cite{zhu2024bicoherence} conducted a bicoherence analysis on the QPOs of MAXI J1535-571 during its September to October 2017 outburst. They discovered that the bicoherence pattern of type C QPOs initially appears as a web pattern and transitions to a hypotenuse pattern following the emergence of type B QPOs. This change indicates that MAXI J1535-571 is a low-inclination source. 

Our results in Figure~\ref{figure3} and ~\ref{figure5} for the LE and HE energy bands distinctly showcase the evolution of bicoherence, clearly transitioning from the web pattern to the hypotenuse pattern. However, there are subtle differences in the evolution process. In the bicoherence patterns of the LE energy band, the diagonal ($f_1+f_2=f_{\rm QPO}$) bicoherence is not high at the initial stage and sometimes even disappears, presenting a cross pattern in certain observations, such as observation ID P061433800512. As the outburst progresses, the bicoherence value at $f_1=f_2=f_{\rm har}$ gradually weakens, as observed in Figure~\ref{figure3} from observation ID P061433800307 to P061433801403. This is similar to the evolutionary characteristics of MAXI J1535-571 as shown in \cite{zhu2024bicoherence} (see their Figure 4). 

The evolutionary characteristics are even more evident when we examine the results shown in Figure~\ref{figure5}. In all observations we utilized, there is no significant bicoherence at $f_1=f_2=f_{\rm har}$, possibly due to the weakening strength of harmonics in the HE energy band. Throughout the entire outburst, the diagonal structure at $f_1+f_2=f_{\rm QPO}$ in the results displaying bicoherence patterns is consistently prominent.

In our previous description of the results, we provided a qualitative explanation of the partial correlation between the changes in bi-coherence patterns and the flares. However, further analysis suggests that these changes may be more strongly correlated with the QPO frequency. For example, as shown in the Figure~\ref{figure3} for observation ID P061433801403, its observation time occurs during the weakening phase of flare-1, and it exhibits strong bicoherence characteristics at $f_1=f_2=f_{\rm QPO}$.  In the HE energy band, the relationship between the bicoherence pattern and the QPO frequency is more pronounced. When the frequency exceeds 2 Hz, the horizontal and vertical bicoherence features disappeared.  

The relationship between the pattern transition and frequency is detailed in the studies by \cite{arur2019non,arur2020likely}.An increase in the optical depth of the corona could lead to a gradual transition in the bicoherence pattern from a web to a hypotenuse pattern. In this interpretation, the initial ‘web’ pattern observed in the low hard state and the hard intermediate state  evolves as the source moves to the the soft intermediate state. As the QPO frequency increases and the spectrum becomes dominated by the disc blackbody component, the abundance of seed photons available for inverse Compton scattering by the corona increases, resulting in a stronger ‘hypotenuse’ pattern in this state.

The bicoherence patterns transitioned from the initial web pattern to the hypotenuse pattern.  In the study by \cite{arur2020likely}, low inclination sources, ranging broadly between 20° and 55°,  exhibited the characteristic transition from a web pattern to a hypotenuse pattern. 
This suggests that Swift J1727.8-1613 is likely a low inclination source with an inclination range of approximately 20 to 55°. This conclusion is consistent with the results obtained from the previous analysis. 

Based on the similarity of the detected X-ray power density in Swift J1727.8-1613 and Cyg X-1, \cite{veledina2023discovery} inferred that Swift J1727.8-1613 is a source with an intermediate inclination, approximately i $\sim$30°-60°. \cite{peng2024nicer} utilized observations from Insight-HXMT, NICER, and NuSTAR during the source's transition to the HIMS, measured an inclination angle of approximately $40_{-0.8}^{+1.2}$°.  This result is consistent with the findings of \cite{draghis2023preliminary}, who also utilized NICER data and arrived at a similar conclusion. \cite{svoboda2024dramatic} presented IXPE results obtained during the HSS. The combined polarization measurements from both the soft and hard states suggest that a very high or low inclination of the system is unlikely and they provided an inclination range of approximately 30-50°.  
  

\subsection{Energy dependence}
In the study by \cite{zhu2024bicoherence}, the relationship between bicoherence and energy is discussed. Based on NICER observations, the energy bands were divided into 1-3 keV and 3-10 keV. The PDS does not show significant differences across these energy bands. However, the bicoherence results indicate that intensity may be correlated with energy bands. The horizontal and vertical lines in the web pattern are more pronounced in the lower energy bands before the appearance of type-B QPOs, while the diagonal lines in the hypotenuse pattern are more prominent in the higher energy bands after type-B QPOs appear (see their Figure 6). 

Thanks to Insight-HXMT's broad energy range, we were able to conduct an in-depth study of the relationship between bicoherence and energy. From our results, depicted in Figures~\ref{figure3} to ~\ref{figure5}, it is evident that the bicoherence patterns exhibit distinct characteristics across different energy bands from 1 -- 100 keV. Using observation ID P061433800307 as an example (more detailed examples can be found in the Appendix~\ref{secapp}), the horizontal and vertical lines in the 'web' pattern are most prominent in the LE band. In the ME band, the results show that the horizontal and vertical lines are very weak.

In the HE band, the horizontal and vertical lines become noticeable; however, the bicoherence at $f_1=f_2=f_{\rm har}$ disappears. In other words, the bicoherence at $f_1=f_2=f_{\rm har}$ gradually weakens and eventually disappears as the energy increases. This is primarily due to the weakening of harmonics at higher energies. In the study of QPOs and their harmonics in GX 339-4 using frequency-resolved spectroscopy,  \cite{axelsson2016revealing} found that in the hard state, the harmonic is stronger than the QPO, while in the softer observations, the harmonic is weaker and its spectrum is very different from either the average spectrum or the QPO. Additionally, at high energies, the harmonic disappears in the power spectra above 10 keV.

For the diagonal structures, they are the least obvious in the LE band. In the ME band, bicoherence not only appears at 
$f_1+f_2=f_{\rm QPO}$ 
  but also shows significant coherence at 
$f_1+f_2=f_{\rm har}$ 
 , forming a 'parallel' pattern. The diagonal structures at $f_1+f_2=f_{\rm QPO}$  are the strongest and most prominent in the  HE band, but the diagonal structure at 
$f_1+f_2=f_{\rm har}$ disappears in the HE band.

We examined the potential impact of count rates. The count rates in the LE, ME, and HE bands are $5569.95 \pm 842.88$, $1484.71 \pm 530.03$, and $3652.88 \pm 709.48$, respectively for P061433800307. From LE to ME and then to HE, the count rates initially decrease and then increase. 
This trend is also evident from the light curve on the left-hand pane of Figure~\ref{figure1}. Similarly, the horizontal and vertical lines at $f_1=f_2=f_{\rm QPO}$ first weaken and then become stronger. However, the bicoherence at $f_1=f_2=f_{\rm har}$ weakens from the LE to ME bands and then disappears in the HE band, showing no clear relationship with the count rates. The diagonal structures at 
$f_1+f_2=f_{\rm QPO}$ 
  gradually strengthen, while the count rates initially decrease and then increase, showing no clear correlation. 
As the count rate first decreases and then increases, the diagonal structures at 
$f_1+f_2=f_{\rm har}$ 
  appear only in the ME energy band.
Count rates may be related to the bicoherence of the horizontal and vertical lines at $f_1=f_2=f_{\rm QPO}$. 
but are not related to the bicoherence of the diagonal lines.  

\subsection{Broad-band noise coupling}
The bicoherence patterns in the LE and HE bands are similar to those reported in previous studies on other sources \citep{arur2019non,arur2020likely,arur2022using,zhu2024bicoherence}. However, a new bicoherence pattern emerges in the ME band. 
\cite{maccarone2011coupling} studied Rossi X-ray Timing Explorer observations of GRS 1915+105 when the source exhibited strong QPOs and strong BBN components in its PDS. 
The results show that the QPO is coupled with the noise components rather than being generated independently. The study explored various mathematical models of light curves reflecting physical scenarios of correlated QPOs and noise components.

In their simulations \citep{maccarone2011coupling}, the output light curve is determined by using two power spectra:noise component and QPO component. The time series for these components are generated under the assumption of a Gaussian process. This ensures that any non-linearity detected in the simulations arises from the physical processes incorporated after the initial random time series generation. A reservoir model (see \citealt{maccarone2011coupling} for mathematical details ) is used where energy is injected either at a constant rate or at a random rate. The reservoir supplies energy to the two components. The flux of each component is calculated as the product of its time series value, the size of the reservoir at that time, and a normalization factor. In this model, if a component drains energy without contributing to the X-ray light curve, the effective size of the reservoir for the X-ray light curve is reduced by the amount of energy drained by the non-contributing component. This modulation of the reservoir size affects the output light curve when multiplied by the QPO time series, leading to a modulation of the QPO on the noise time-scale and altering the final power spectrum. When the amplitude of variability is large, the reservoir's strong fluctuations produce QPO harmonics and phase coupling with noise, leading to a web-like pattern in the bicoherence. However, this simulated pattern differs from the real data, where cross-like structures only appear at higher noise frequencies. In low variability amplitude cases, the simulated bicoherence shows prominent cross-like structures and a weak hypotenuse when noise is added. 

Upon comparing our observed bicoherence results in the ME band with the simulated outcomes in \cite{maccarone2011coupling}, we found consistency with their findings presented in Figure 3(a). 
This pattern generated in these simulations occurs under conditions of low variability amplitude and without the noise component added back in. If the noise component is added back in, the bicoherence pattern will transform into a web pattern, as illustrated in Figure 4(a) in \cite{maccarone2011coupling}. 
This means that the bicoherence pattern observed in the ME energy band should correspond to the same conditions replicated in the reservoir model \citep{maccarone2011coupling}: low amplitude variability with minimal coupling of the noise component. 
\section{CONCLUSIONS}
\label{conclusion}
Based on the Insight-HXMT observational data of Swift J1727.8-1613, we employed the bicoherence method to analyze Type C QPOs with frequencies greater than $\sim$ 1 Hz. The main results of the study can be summarized as follows:
\begin{itemize}
\item The frequency of type C QPOs increased over time. During the flare-1 and flare-2 phases, there was a sudden increase in the QPO frequency, accompanied by a significant decrease in the hardness ratio. There is a strong correlation between the QPO frequency, QPO RMS, and hardness ratio. We used power functions and linear functions to fit these correlations separately, providing the parameter-fitting results. 
\item 
Using the bicoherence method, we observed a transition from web patterns to hypotenuse patterns in the LE and HE results. According to the relationship between bicoherence patterns and inclination, this suggests that Swift J1727.8-1613 is likely a low-inclination ($\sim$ 30-50°) source. 
\item Analysis of the bicoherence results for different energy bands and varying count rates suggests that the bicoherence of the horizontal and vertical lines at  $f_1=f_2=f_{\rm har}$ appears to be more dependent on energy. Additionally, the diagonal structure at  $f_1+f_2=f_{\rm QPO}$ shows a characteristic strengthening with increasing energy. 

\item In the ME band, we have for the first time observed the 'parallel' bicoherence patterns in the observational data. The emergence of these patterns may be related to the coupling properties between the BBN and various components. 

\end{itemize}

\section*{Acknowledgements}

We thank the anonymous referee for insightful comments and useful
suggestions that improved the paper. This work is supported by the National Key Research and Development Program of China (Grants No. 2021YFA0718503 and 2023YFA1607901), the NSFC (12133007). This work has made use of data from the \textit{Insight-}HXMT mission, a project funded by the China National Space Administration (CNSA) and the Chinese Academy of Sciences (CAS).


\bibliography{sample631}{}

\begin{thebibliography}{}
\expandafter\ifx\csname natexlab\endcsname\relax\def\natexlab#1{#1}\fi
\providecommand{\url}[1]{\href{#1}{#1}}
\providecommand{\dodoi}[1]{doi:~\href{http://doi.org/#1}{\nolinkurl{#1}}}
\providecommand{\doeprint}[1]{\href{http://ascl.net/#1}{\nolinkurl{http://ascl.net/#1}}}
\providecommand{\doarXiv}[1]{\href{https://arxiv.org/abs/#1}{\nolinkurl{https://arxiv.org/abs/#1}}}

\bibitem[{Arur \& Maccarone(2019)}]{arur2019non}
Arur, K., \& Maccarone, T. 2019, Monthly Notices of the Royal Astronomical
  Society, 486, 3451

\bibitem[{Arur \& Maccarone(2020)}]{arur2020likely}
---. 2020, Monthly Notices of the Royal Astronomical Society, 491, 313

\bibitem[{Arur \& Maccarone(2022)}]{arur2022using}
Arur, K., \& Maccarone, T.~J. 2022, Monthly Notices of the Royal Astronomical
  Society, 514, 1720

\bibitem[{Axelsson \& Done(2016)}]{axelsson2016revealing}
Axelsson, M., \& Done, C. 2016, Monthly Notices of the Royal Astronomical
  Society, 458, 1778

\bibitem[{Belloni \& Hasinger(1990)}]{belloni1990atlas}
Belloni, T., \& Hasinger, G. 1990, Astronomy and Astrophysics, 230, 103

\bibitem[{Bollemeijer {et~al.}(2023)Bollemeijer, Uttley, Buisson, Homan,
  Altamirano, Gendreau, Arzoumanian, Strohmayer, \&
  Sanna}]{bollemeijer2023nicer}
Bollemeijer, N., Uttley, P., Buisson, D., {et~al.} 2023, ATel, 16247, 1

\bibitem[{Bu {et~al.}(2015)Bu, Chen, Li, Qu, Belloni, \&
  Zhang}]{bu2015correlations}
Bu, Q.-c., Chen, L., Li, Z.-s., {et~al.} 2015, The Astrophysical Journal, 799,
  2

\bibitem[{Cao {et~al.}(2020)Cao, Jiang, Meng, Zhang, Luo, Yang, Zhang, Gu, Sun,
  Liu, {et~al.}}]{cao2020medium}
Cao, X., Jiang, W., Meng, B., {et~al.} 2020, Science China Physics, Mechanics
  \& Astronomy, 63, 1

\bibitem[{Casella {et~al.}(2005)Casella, Belloni, \& Stella}]{casella2005abc}
Casella, P., Belloni, T., \& Stella, L. 2005, The Astrophysical Journal, 629,
  403

\bibitem[{Castro-Tirado {et~al.}(2023)Castro-Tirado, Sanchez-Ramirez,
  Caballero-Garcia, Perez-Garcia, Fernandez-Garcia, Guziy, Hu, Blazek, Hermelo,
  Pinter, {et~al.}}]{castro2023optical}
Castro-Tirado, A., Sanchez-Ramirez, R., Caballero-Garcia, M., {et~al.} 2023,
  ATel, 16208, 1

\bibitem[{Chakrabarti {et~al.}(2008)Chakrabarti, Debnath, Nandi, \&
  Pal}]{chakrabarti2008evolution}
Chakrabarti, S.~K., Debnath, D., Nandi, A., \& Pal, P. 2008, \aap, 489, L41

\bibitem[{{Chakrabarti} \& {Molteni}(1993)}]{chakrabarti1993smoothed}
{Chakrabarti}, S.~K., \& {Molteni}, D. 1993, \apj, 417, 671

\bibitem[{Chatterjee {et~al.}(2024)Chatterjee, Mondal, Singh, \&
  Sugizaki}]{chatterjee2024insight}
Chatterjee, K., Mondal, S., Singh, C.~B., \& Sugizaki, M. 2024, arXiv preprint
  arXiv:2405.01498

\bibitem[{Chen \& Wang(2024)}]{chen2024different}
Chen, X., \& Wang, W. 2024, Journal of High Energy Astrophysics, 41, 89

\bibitem[{Chen {et~al.}(2022{\natexlab{a}})Chen, Wang, You, Tian, Liu, Zhang,
  Ding, Qu, Zhang, Song, {et~al.}}]{chen2022wavelet}
Chen, X., Wang, W., You, B., {et~al.} 2022{\natexlab{a}}, Monthly Notices of
  the Royal Astronomical Society, 513, 4875

\bibitem[{Chen {et~al.}(2022{\natexlab{b}})Chen, Wang, Tian, Zhang, Liu, Wu,
  Sai, Huang, Song, Qu, {et~al.}}]{chen2022waveletB}
Chen, X., Wang, W., Tian, P., {et~al.} 2022{\natexlab{b}}, Monthly Notices of
  the Royal Astronomical Society, 517, 182

\bibitem[{Chen {et~al.}(2020)Chen, Cui, Li, Wang, Xu, Lu, Wang, Chen, Han, Hu,
  {et~al.}}]{chen2020low}
Chen, Y., Cui, W., Li, W., {et~al.} 2020, Science China Physics, Mechanics \&
  Astronomy, 63, 1

\bibitem[{{Dichiara} {et~al.}(2023){Dichiara}, {Kennea}, {Page}, {Parsotan},
  {Williams}, \& {Neil Gehrels Swift Observatory Team}}]{Kennea2023GCN}
{Dichiara}, S., {Kennea}, J.~A., {Page}, K.~L., {et~al.} 2023, GCN, 34542, 1

\bibitem[{Ding {et~al.}(2023)Ding, Ji, Bu, Dong, \& Chang}]{ding2023nonlinear}
Ding, Q., Ji, L., Bu, Q.-C., Dong, T., \& Chang, J. 2023, Research in Astronomy
  and Astrophysics, 23, 085024

\bibitem[{Dovciak {et~al.}(2023{\natexlab{a}})Dovciak, Ratheesh, Tennant, \&
  Ma}]{dovciak2023ixpe}
Dovciak, M., Ratheesh, A., Tennant, A., \& Ma, G. 2023{\natexlab{a}}, ATel,
  16237, 1

\bibitem[{Dovciak {et~al.}(2023{\natexlab{b}})Dovciak, Ratheesh, Tennant, \&
  Ma}]{dovciak2023ixpeb}
---. 2023{\natexlab{b}}, ATel, 16242, 1

\bibitem[{Draghis {et~al.}(2023)Draghis, Miller, Homan, Uttley, Bollemeijer,
  Steiner, Hare, Tombesi, Gendreau, Arzoumanian,
  {et~al.}}]{draghis2023preliminary}
Draghis, P.~A., Miller, J.~M., Homan, J., {et~al.} 2023, The Astronomer's
  Telegram, 16219, 1

\bibitem[{Hamilton(2020)}]{hamilton2020time}
Hamilton, J.~D. 2020, Time series analysis (Princeton university press)

\bibitem[{Huang {et~al.}(1998)Huang, Shen, Long, Wu, Shih, Zheng, Yen, Tung, \&
  Liu}]{Huang1998}
Huang, N.~E., Shen, Z., Long, S.~R., {et~al.} 1998, Proceedings of the Royal
  Society of London. Series A: Mathematical, Physical and Engineering Sciences,
  454, 903, \dodoi{10.1098/rspa.1998.0193}

\bibitem[{Ingram \& Done(2011)}]{ingram2011physical}
Ingram, A., \& Done, C. 2011, Monthly Notices of the Royal Astronomical
  Society, 415, 2323

\bibitem[{Ingram {et~al.}(2009)Ingram, Done, \& Fragile}]{ingram2009low}
Ingram, A., Done, C., \& Fragile, P.~C. 2009, Monthly Notices of the Royal
  Astronomical Society: Letters, 397, L101

\bibitem[{Ingram {et~al.}(2023)Ingram, Bollemeijer, Veledina, Dovciak,
  Poutanen, Egron, Russell, Trushkin, Negro, Ratheesh,
  {et~al.}}]{ingram2023tracking}
Ingram, A., Bollemeijer, N., Veledina, A., {et~al.} 2023, arXiv preprint
  arXiv:2311.05497

\bibitem[{Ingram(2019)}]{ingram2019}
Ingram, A.~R. 2019, New Astronomy Reviews, 29

\bibitem[{Katoch {et~al.}(2023)Katoch, Antia, Nandi, \&
  Shah}]{katoch2023detection}
Katoch, T., Antia, H., Nandi, A., \& Shah, P. 2023, ATel, 16235, 1

\bibitem[{Kim \& Powers(1979)}]{kim1979digital}
Kim, Y.~C., \& Powers, E.~J. 1979, IEEE transactions on plasma science, 7, 120

\bibitem[{Liu {et~al.}(2020)Liu, Zhang, Li, Lu, Chang, Li, Zhang, Jin, Yu,
  Zhang, {et~al.}}]{liu2020High}
Liu, C., Zhang, Y., Li, X., {et~al.} 2020, Science China Physics, Mechanics \&
  Astronomy, 63, 1

\bibitem[{Ma {et~al.}(2021)Ma, Tao, Zhang, Zhang, Bu, Ge, Chen, Qu, Zhang, Lu,
  {et~al.}}]{ma2021discovery}
Ma, X., Tao, L., Zhang, S.-N., {et~al.} 2021, Nature Astronomy, 5, 94

\bibitem[{Ma {et~al.}(2023)Ma, Zhang, Tao, Bu, Qu, Zhang, Zhou, Huang, Jia,
  Song, {et~al.}}]{ma2023detailed}
Ma, X., Zhang, L., Tao, L., {et~al.} 2023, arXiv preprint arXiv:2303.00481

\bibitem[{Maccarone(2013)}]{maccarone2013biphase}
Maccarone, T.~J. 2013, Monthly Notices of the Royal Astronomical Society, 435,
  3547

\bibitem[{Maccarone \& Coppi(2002)}]{maccarone2002higher}
Maccarone, T.~J., \& Coppi, P.~S. 2002, Monthly Notices of the Royal
  Astronomical Society, 336, 817

\bibitem[{Maccarone {et~al.}(2011)Maccarone, Uttley, Van Der~Klis, Wijnands, \&
  Coppi}]{maccarone2011coupling}
Maccarone, T.~J., Uttley, P., Van Der~Klis, M., Wijnands, R.~A., \& Coppi,
  P.~S. 2011, Monthly Notices of the Royal Astronomical Society, 413, 1819

\bibitem[{Mereminskiy {et~al.}(2023)Mereminskiy, Lutovinov, Molkov, Krivonos,
  Semena, Sazonov, Tkachenko, \& Sunyaev}]{mereminskiy2023hard}
Mereminskiy, I., Lutovinov, A., Molkov, S., {et~al.} 2023, arXiv preprint
  arXiv:2310.06697

\bibitem[{Miller-Jones {et~al.}(2023)Miller-Jones, Sivakoff, Bahramian, \&
  Russell}]{miller2023vla}
Miller-Jones, J., Sivakoff, G., Bahramian, A., \& Russell, T. 2023, ATel,
  16211, 1

\bibitem[{Miyamoto {et~al.}(1991)Miyamoto, Kimura, Kitamoto, Dotani, \&
  Ebisawa}]{miyamoto1991x}
Miyamoto, S., Kimura, K., Kitamoto, S., Dotani, T., \& Ebisawa, K. 1991, The
  Astrophysical Journal, 383, 784

\bibitem[{{Molteni} {et~al.}(1996){Molteni}, {Sponholz}, \&
  {Chakrabarti}}]{Molteni1996}
{Molteni}, D., {Sponholz}, H., \& {Chakrabarti}, S.~K. 1996, \apj, 457, 805

\bibitem[{Motta {et~al.}(2015)Motta, Casella, Henze, Mu{\~n}oz-Darias, Sanna,
  Fender, \& Belloni}]{motta2015geometrical}
Motta, S., Casella, P., Henze, M., {et~al.} 2015, Monthly Notices of the Royal
  Astronomical Society, 447, 2059

\bibitem[{Motta(2016)}]{motta2016quasi}
Motta, S.~E. 2016, Astronomische Nachrichten, 337, 398

\bibitem[{Negoro {et~al.}(2023)Negoro, Serino, Nakajima, Kobayashi, Tanaka,
  Soejima, Kudo, Mihara, Kawamuro, Yamada, {et~al.}}]{negoro2023maxi}
Negoro, H., Serino, M., Nakajima, M., {et~al.} 2023, ATel, 16205, 1

\bibitem[{O'Connor {et~al.}(2023)O'Connor, Hare, Younes, Gendreau, Arzoumanian,
  \& Ferrara}]{o2023nicer}
O'Connor, B., Hare, J., Younes, G., {et~al.} 2023, GCN, 34549, 1

\bibitem[{{Page} {et~al.}(2023){Page}, {Dichiara}, {Gropp}, {Krimm},
  {Parsotan}, {Williams}, \& {Neil Gehrels Swift Observatory Team}}]{Page2023}
{Page}, K.~L., {Dichiara}, S., {Gropp}, J.~D., {et~al.} 2023, GCN, 34537, 1

\bibitem[{Palmer \& Parsotan(2023)}]{palmer2023swift}
Palmer, D.~M., \& Parsotan, T.~M. 2023, ATel, 16215, 1

\bibitem[{Peng {et~al.}(2024)Peng, Zhang, Shui, Zhang, Kong, Chen, Wang, Ji,
  Qu, Tao, {et~al.}}]{peng2024nicer}
Peng, J.-Q., Zhang, S., Shui, Q.-C., {et~al.} 2024, The Astrophysical Journal
  Letters, 960, L17

\bibitem[{Remillard \& McClintock(2006)}]{remillard2006x}
Remillard, R.~A., \& McClintock, J.~E. 2006, Annu. Rev. Astron. Astrophys., 44,
  49

\bibitem[{S{\'a}nchez {et~al.}(2024)S{\'a}nchez, Mu{\~n}oz-Darias, Padilla,
  Casares, \& Torres}]{sanchez2024evidence}
S{\'a}nchez, D.~M., Mu{\~n}oz-Darias, T., Padilla, M.~A., Casares, J., \&
  Torres, M. 2024, \aap, 682, L1

\bibitem[{Schnittman {et~al.}(2006)Schnittman, Homan, \&
  Miller}]{schnittman2006precessing}
Schnittman, J.~D., Homan, J., \& Miller, J.~M. 2006, The Astrophysical Journal,
  642, 420

\bibitem[{Shui {et~al.}(2024)Shui, Zhang, Zhang, Chen, Kong, Peng, Ji, Wang,
  Chang, Zhuo, {et~al.}}]{shui2024recovery}
Shui, Q.~C., Zhang, S., Zhang, S.~N., {et~al.} 2024, The Astrophysical Journal
  Letters, 965, L7

\bibitem[{Stella \& Vietri(1997)}]{stella1997lense}
Stella, L., \& Vietri, M. 1997, The Astrophysical Journal, 492, L59

\bibitem[{Stella {et~al.}(1999)Stella, Vietri, \&
  Morsink}]{stella1999correlations}
Stella, L., Vietri, M., \& Morsink, S.~M. 1999, The Astrophysical Journal, 524,
  L63

\bibitem[{Sunyaev {et~al.}(2023)Sunyaev, Mereminskiy, Molkov, Semena, Arefiev,
  Krivonos, Levin, Lutovinov, Shtykovsky, \& Tkachenko}]{sunyaev2023integral}
Sunyaev, R., Mereminskiy, I., Molkov, S., {et~al.} 2023, ATel, 16217, 1

\bibitem[{Svoboda {et~al.}(2024)Svoboda, Dov{\v{c}}iak, Steiner, Kaaret,
  Podgorn{\`y}, Poutanen, Veledina, Muleri, Taverna, Krawczynski,
  {et~al.}}]{svoboda2024dramatic}
Svoboda, J., Dov{\v{c}}iak, M., Steiner, J.~F., {et~al.} 2024, The
  Astrophysical Journal Letters, 966, L35

\bibitem[{{Tagger} \& {Pellat}(1999)}]{Tagger1999}
{Tagger}, M., \& {Pellat}, R. 1999, \aap, 349, 1003

\bibitem[{Van~der Klis(1989)}]{van1989quasi}
Van~der Klis, M. 1989, ARA\& A, 27, 517

\bibitem[{Varniere \& Tagger(2002)}]{varniere2002accretion}
Varniere, P., \& Tagger, M. 2002, \aap, 394, 329

\bibitem[{Veledina {et~al.}(2023)Veledina, Muleri, Dov{\v{c}}iak, Poutanen,
  Ratheesh, Capitanio, Matt, Soffitta, Tennant, Negro,
  {et~al.}}]{veledina2023discovery}
Veledina, A., Muleri, F., Dov{\v{c}}iak, M., {et~al.} 2023, ApJL, 958, L16

\bibitem[{Wijnands \& van~der Klis(1999)}]{wijnands1999broadband}
Wijnands, R., \& van~der Klis, M. 1999, The Astrophysical Journal, 514, 939

\bibitem[{Wood {et~al.}(2024)Wood, Miller-Jones, Bahramian, Tingay, Prabu,
  Russell, Atri, Carotenuto, Altamirano, Motta, {et~al.}}]{wood2024swift}
Wood, C.~M., Miller-Jones, J.~C., Bahramian, A., {et~al.} 2024, arXiv preprint
  arXiv:2405.12370

\bibitem[{Yu {et~al.}(2023)Yu, Bu, Yang, Liu, Zhang, Huang, Zhou, Qu, Zhang,
  Zhang, {et~al.}}]{yu2023hilbert}
Yu, W., Bu, Q.-C., Yang, Z.-X., {et~al.} 2023, The Astrophysical Journal, 951,
  130

\bibitem[{Yu {et~al.}(2024)Yu, Bu, Zhang, Liu, Zhang, Ducci, Tao, Santangelo,
  Doroshenko, Huang, {et~al.}}]{yu2024timing}
Yu, W., Bu, Q.-C., Zhang, S.-N., {et~al.} 2024, Monthly Notices of the Royal
  Astronomical Society, 529, 4624

\bibitem[{Zhao {et~al.}(2024)Zhao, Tao, Li, Zhang, Feng, Ge, Ji, Wang, Huang,
  Ma, {et~al.}}]{zhao2024first}
Zhao, Q.-C., Tao, L., Li, H.-C., {et~al.} 2024, ApJL, 961, L42

\bibitem[{Zhu \& Wang(2024)}]{zhu2024energy}
Zhu, H., \& Wang, W. 2024, The Astrophysical Journal, 968, 106

\bibitem[{Zhu {et~al.}(2024)Zhu, Chen, \& Wang}]{zhu2024bicoherence}
Zhu, Z., Chen, X., \& Wang, W. 2024, Monthly Notices of the Royal Astronomical
  Society, 529, 4602

\end{thebibliography}
\bibliographystyle{aasjournal}

\clearpage
\appendix
\renewcommand*\thetable{\Alph{section}.\arabic{table}}
\renewcommand*\thefigure{\Alph{section}\arabic{figure}}
\section{Energy dependence of bicoherence}
\label{secapp}
\setcounter{figure}{0}
\begin{figure*}[h]
   \center
    \includegraphics[scale=0.3]{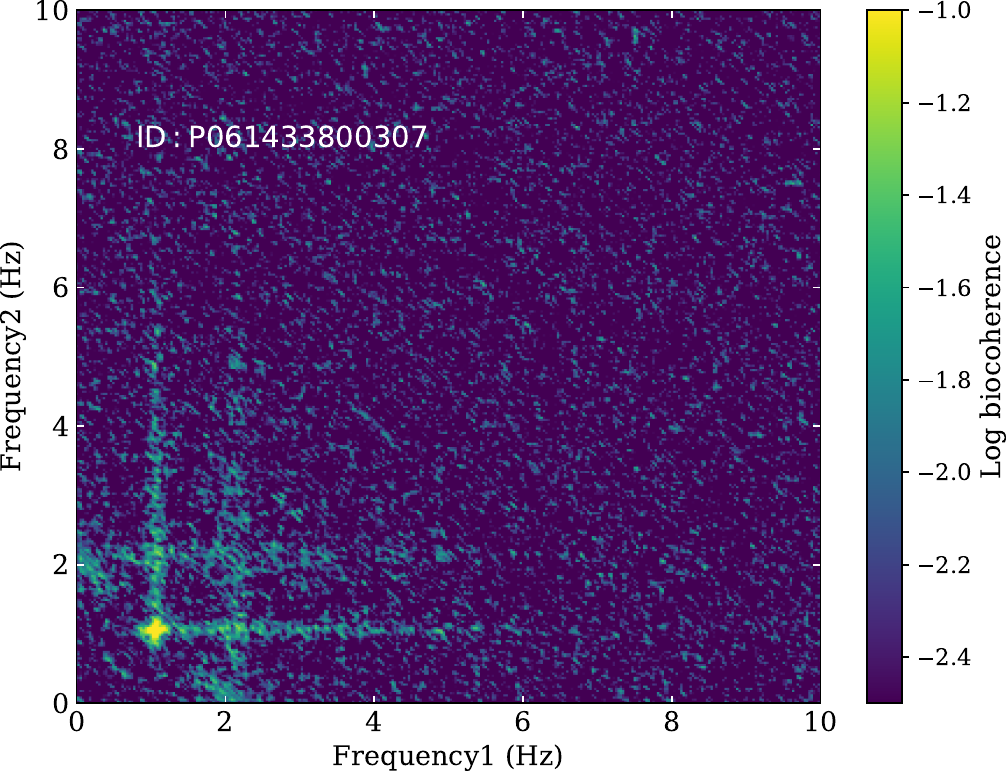}
    \includegraphics[scale=0.3]{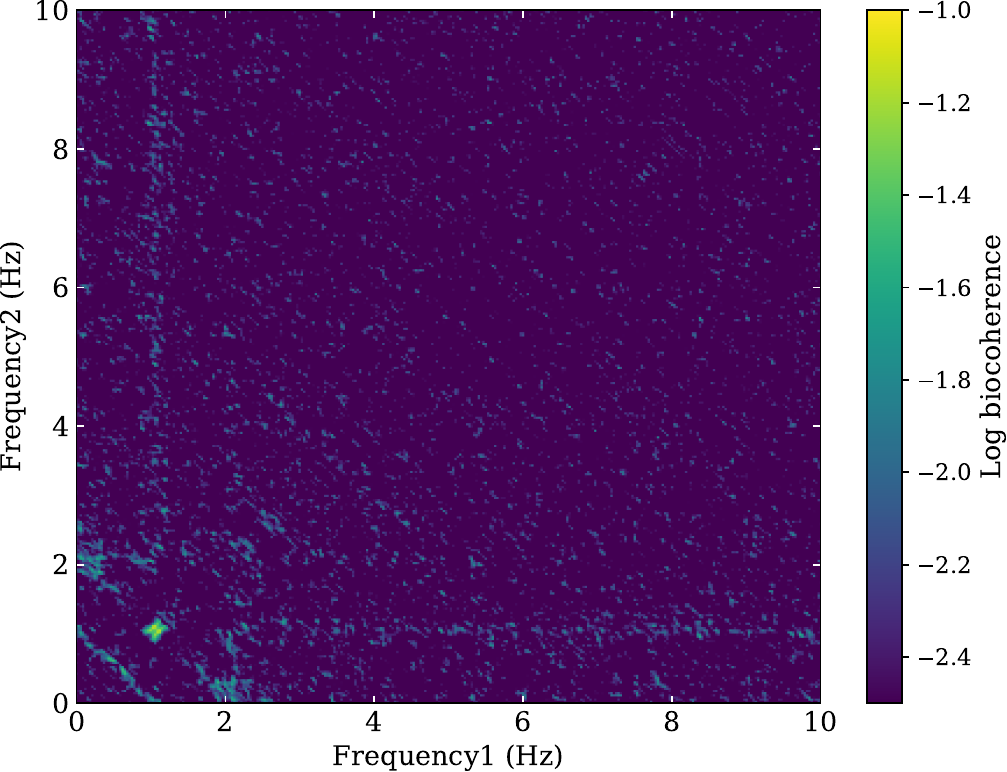}
    \includegraphics[scale=0.3]{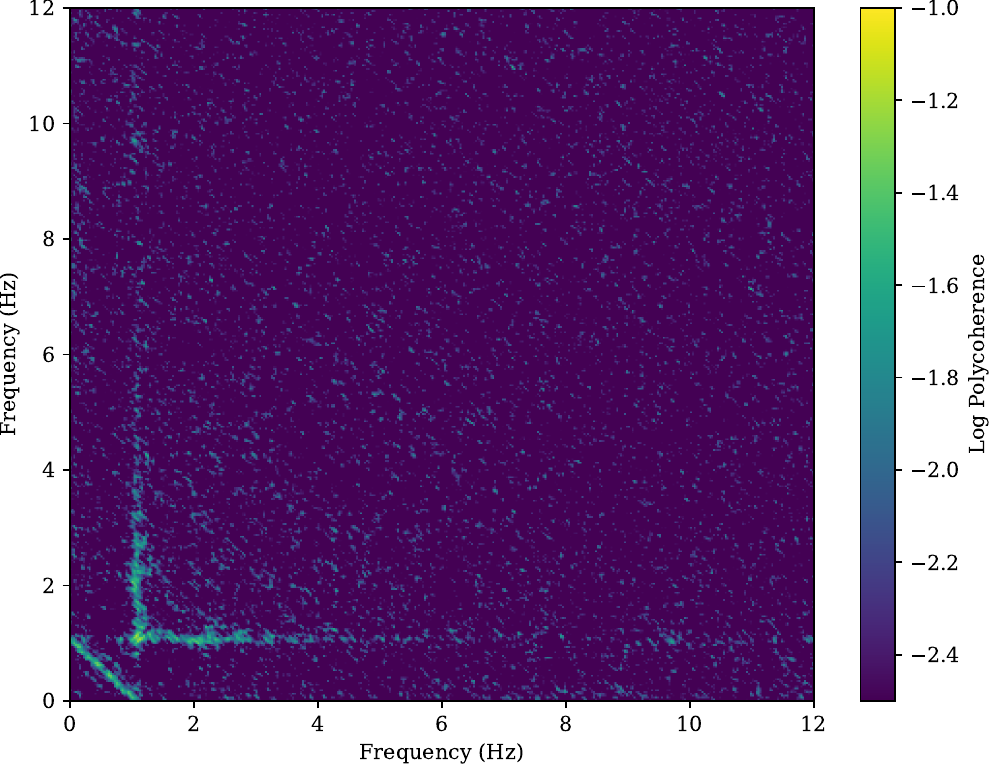}\\
    \includegraphics[scale=0.3]{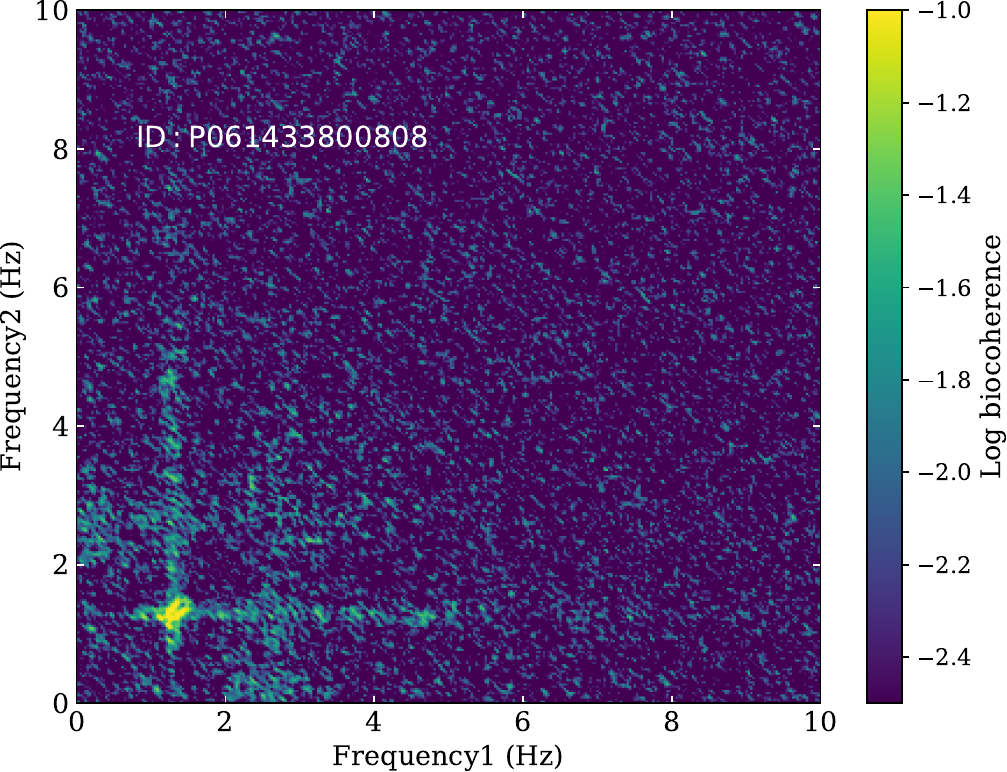}
    \includegraphics[scale=0.3]{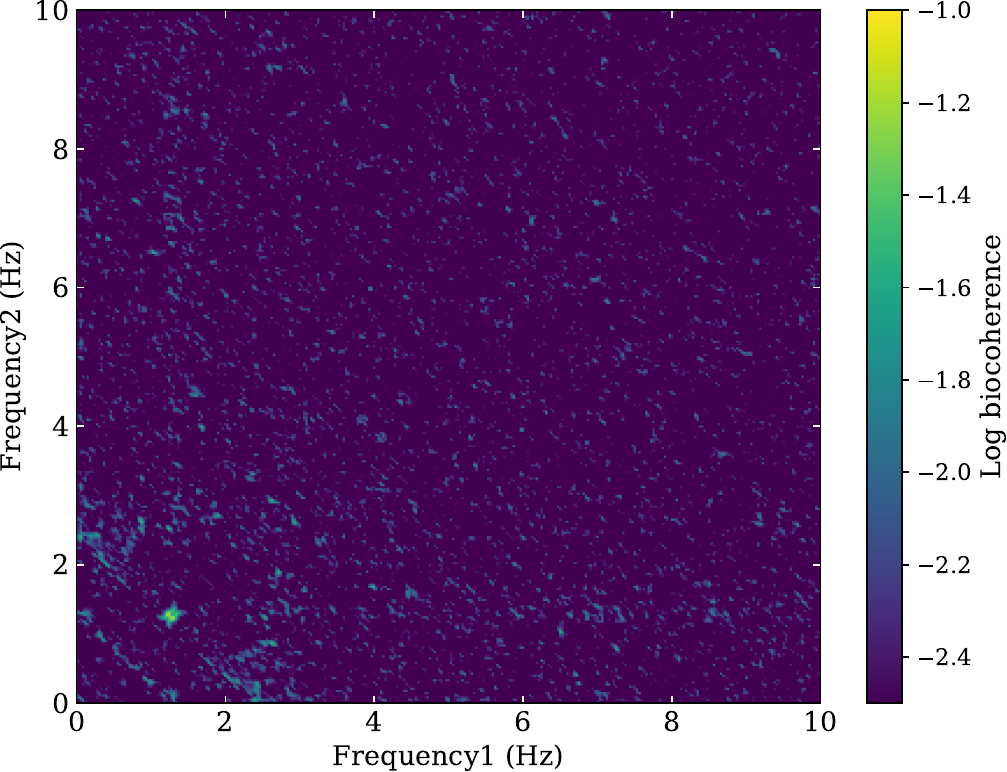}
    \includegraphics[scale=0.3]{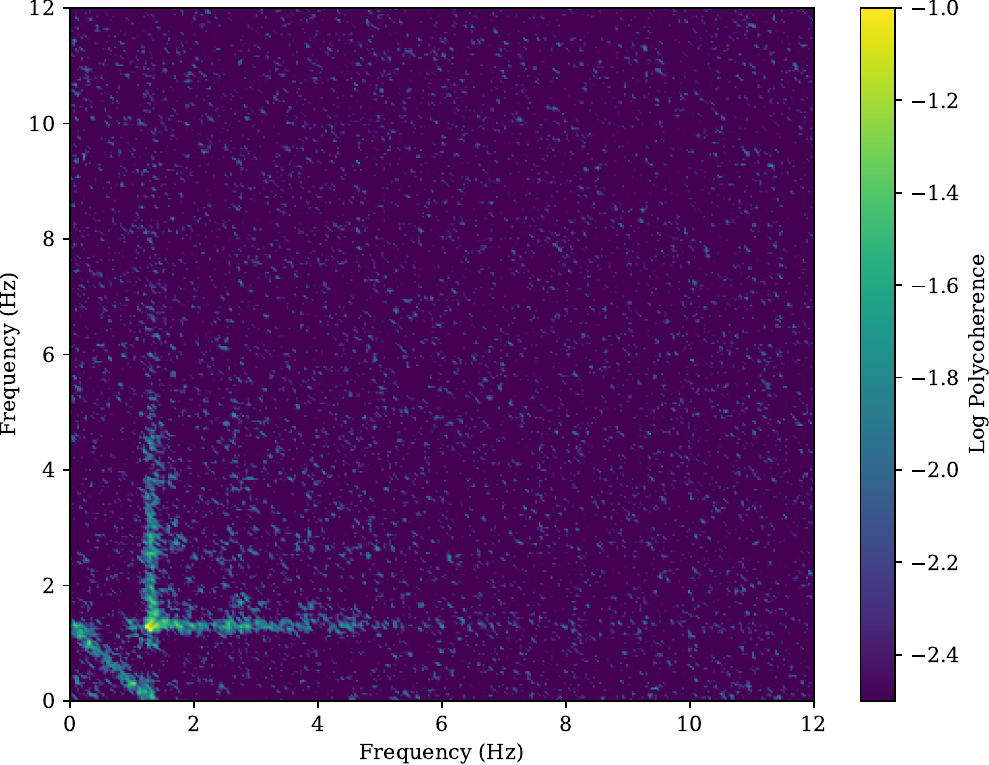}\\
    \includegraphics[scale=0.3]{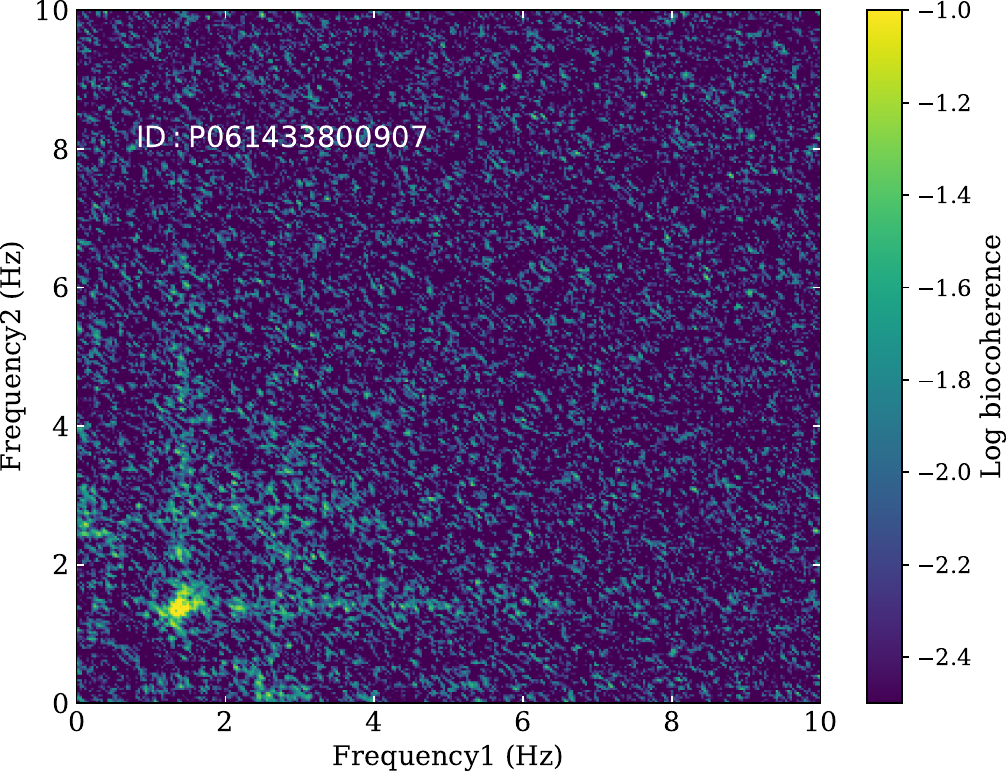}
    \includegraphics[scale=0.3]{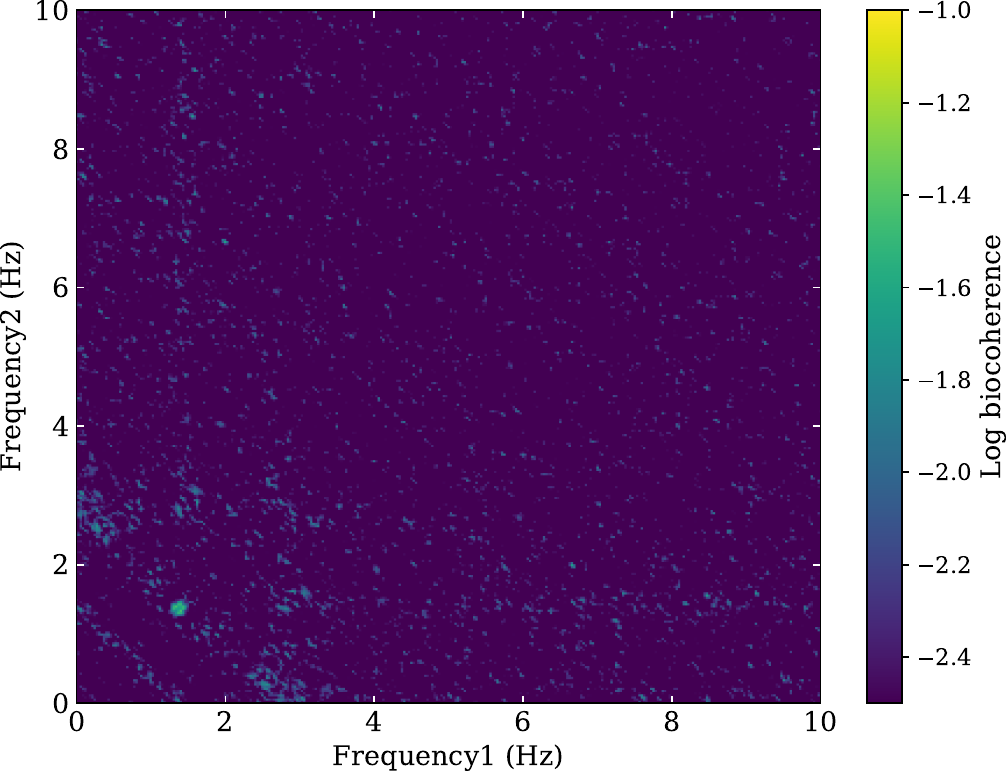}
    \includegraphics[scale=0.3]{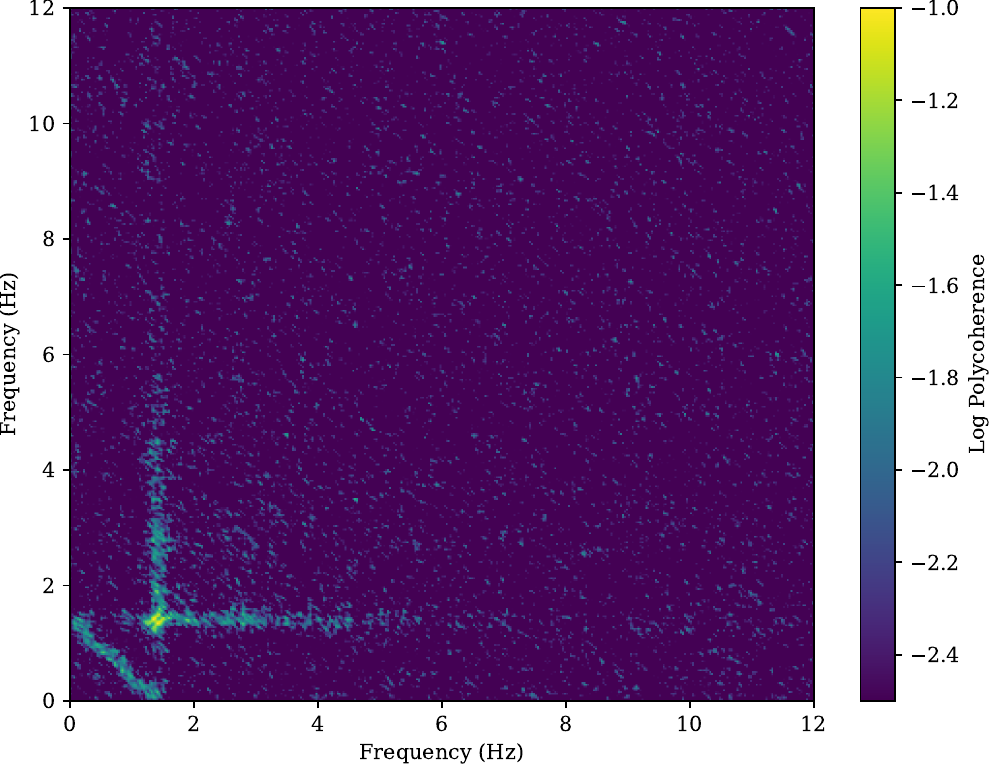}\\
    \includegraphics[scale=0.3]{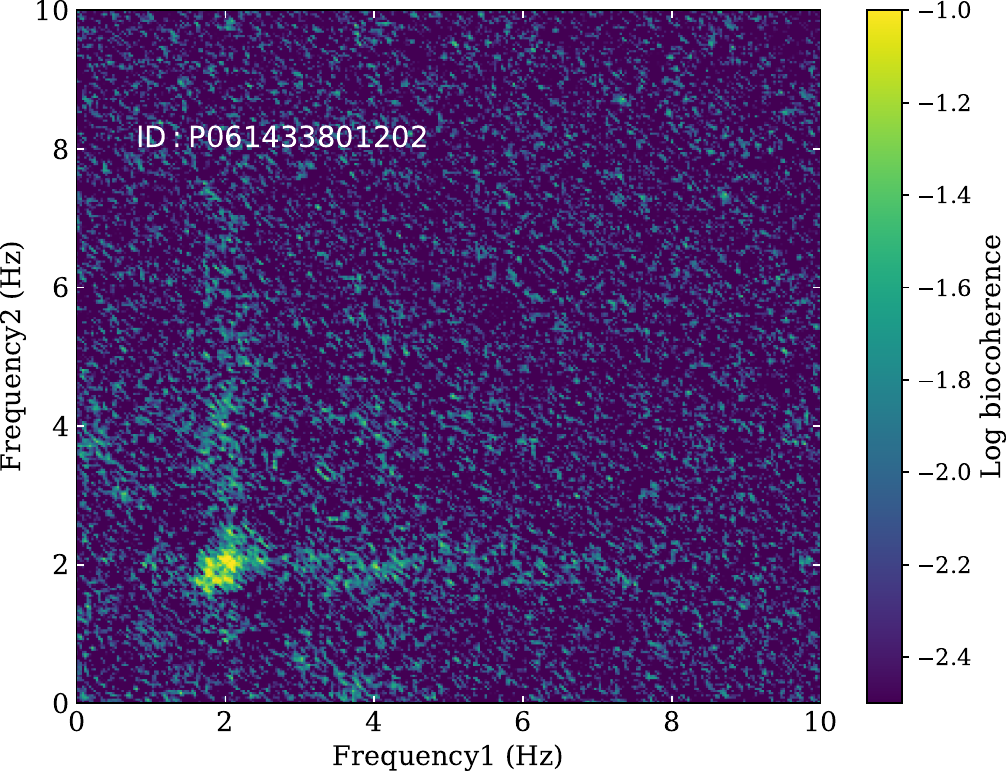}
    \includegraphics[scale=0.3]{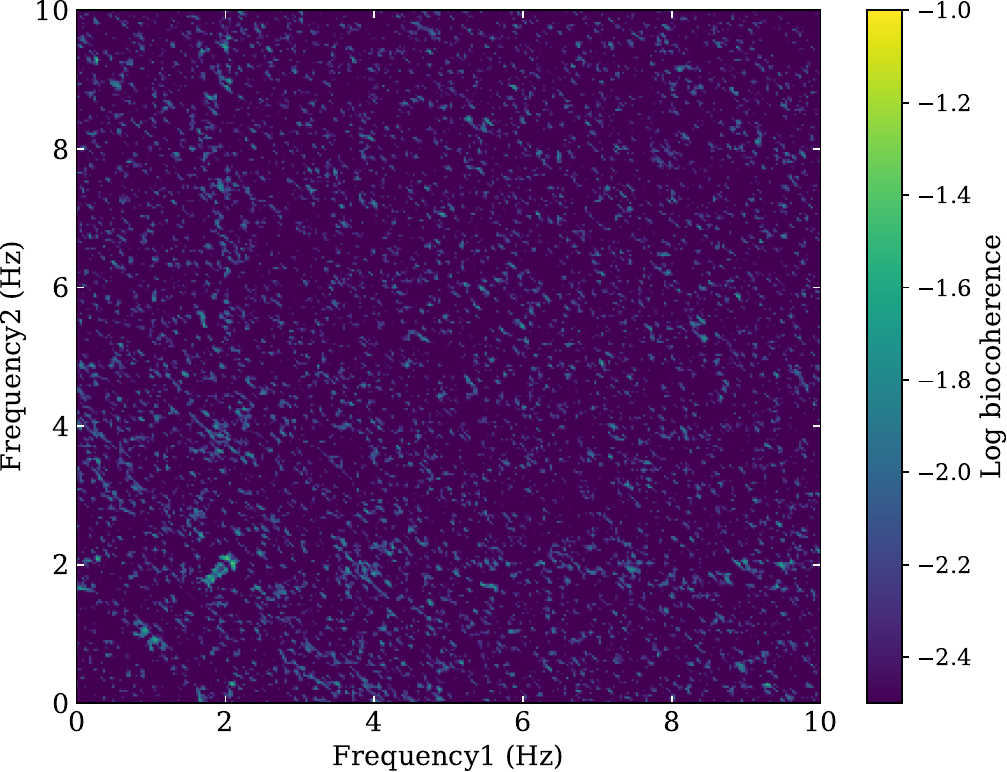}
    \includegraphics[scale=0.3]{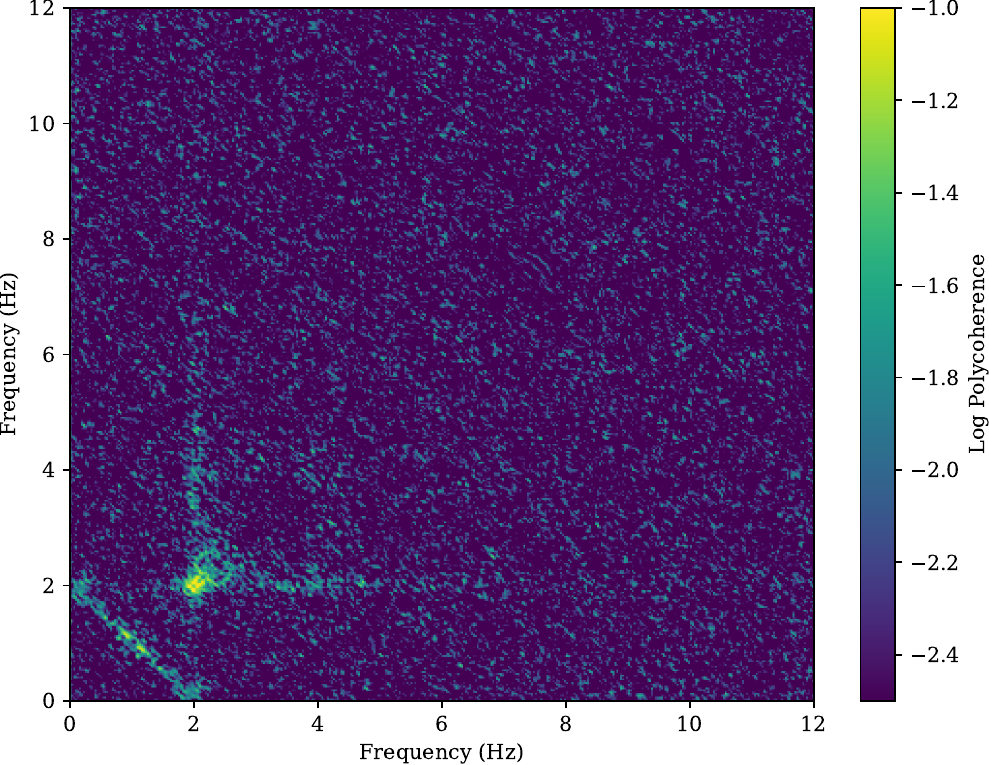}\\

    \caption{The three columns from left to right represent the bicoherence calculation results for LE, ME, and HE, respectively. The corresponding observation IDs are labeled in the first column (LE) of the results. The observation IDs in the second and third columns (ME and HE) are identical to those in the first column.
}
    \label{appfig}
\end{figure*}

\end{document}